\documentclass[aps,prl,twocolumn,showpacs,preprintnumbers]{revtex4-1}

\usepackage{psfrag,graphicx}
\usepackage{dcolumn}
\usepackage{amsmath,amssymb}
\usepackage{bm}
\usepackage{amsfonts,amssymb,amsmath}        
\usepackage{epstopdf}
\usepackage{mathtools}

\usepackage{subcaption} 

\newcommand{\be}{\begin{equation}}
\newcommand{\ee}{\end{equation}}
\newcommand{\bq}{\begin{eqnarray}}
\newcommand{\eq}{\end{eqnarray}}

\newcommand{\ket}[1]{\left | \, #1 \right\rangle}

\begin{document}
\title{Benchmarking of quantum processors with random circuits}
\author{James R. Wootton}
\affiliation{Department of Physics, University of Basel, Klingelbergstrasse 82, CH-4056 Basel, Switzerland}

\begin{abstract}

Quantum processors with sizes in the 10-100 qubit range are now increasingly common. However, with increased size comes increased complexity for benchmarking. The effectiveness of a given device may vary greatly between different tasks, and will not always be  easy to predict from single and two qubit gate fidelities. For this reason, it is important to assess processor quality for a range of important tasks. In this work we propose and implement tests based on random quantum circuits. These are used to evaluate multiple different superconducting qubit devices, with sizes from 5 to 19 qubits, from two hardware manufacturers: IBM Research and Rigetti. The data is analysed to give a quantitative description of how the devices perform.  We also describe how it can be used for a qualitative description accessible to the layperson, by being played as a game.

\end{abstract}


\maketitle

\section{Introduction}

The state of $n$ bits resides within a space of $2^n$ bit strings. By charting a suitable course through this space, classical computers can solve virtually any problem.

The state of $n$ qubits is described by a $2^n$ dimensional Hilbert space. This more general structure allows a new and more subtle ways to move around the space, giving us new and more efficient routes from input to output. The exponential speedups this will allow for certain problems is the primary motivation behind quantum computation.

To determine whether any given quantum processors can live up to this promise, they need to be benchmarked. This could be done using techniques specifically designed for the task, such as randomized benchmarking \cite{emerson:05,magesan:11} or measuring the quantum volume \cite{moll:17}. It could also be done by performing test instances of important quantum algorithms \cite{coles:18} or quantum error correction~\cite{wootton:18,naveh:18}. Whichever is used, the results will depend greatly on the noise levels of the device and also its size and connectivity. The insights gained will therefore be highly dependent on the details of implementation, with the results from a given instance of a given algorithm not providing an unambiguous predictor for the results of others.

To supplement such results, we can seek a task which is more universal in scope. One that can be implemented on devices of any size and connectivity, which takes up the whole of the device, and which directly tests the most important primitive for quantum computing: the ability to fully explore the multiqubit Hilbert space.

Here we propose such a task based on random quantum circuits~\cite{boixo:18}. By implementing random programs, the resulting output states are random samples from the Hilbert space of the device. For short depth random circuits, this sampling will be of states with short-range entanglement that are close to the product states. But for sufficiently long circuits, which allow for the build-up of entanglement across the device, the states will be sampled uniformly from across the entire Hilbert space. Measuring the qubits will then generate bit strings according to a Porter-Thomas distribution, which provides an observable signature of this quantum chaotic regime.

The main application of such sampling will be to act as a test of computational power. Entering into the Porter-Thomas regime for a sufficiently large quantum device would allow a demonstration of quantum computers outperforming classical computers: a milestone known as \textit{quantum computational supremacy}~\cite{preskill:12}. To acheive this, devices will need to be much larger than those considered in this study \cite{chen:18}. Nevertheless, analysis of random circuits for smaller devices will help benchmark our progress towards this milestone, as well as towards the longer term and more important goal of scalable and fault-tolerant quantum computation.

In this paper we perform benchmarking based on random circuits to protoype quantum devices available on the cloud. Specifically, we use the 5 and 16 qubit devices of IBM \cite{ibm:backends} and the 8 and 19 qubit devices of Rigetti \cite{rigetti:acorn}.

\section{Generation of random circuits}

For any given device, we will have a set of native gates to work with. These will include arbitrary single qubit rotations, and entangling gates. The latter are typically two qubit controlled operations, such as the controlled-NOT or controlled-$Z$. In general, it will not be possible to directly implement these between any given pair of qubits on the device. Instead we will have a connectivity graph, in which qubits are nodes that are connected only by an edge when direct coupling is possible. For most physical implementations of qubits, the most straightforward connectivities to realize are a line, for which qubits can couple directly only to their two neighbours, or a ladder composed of two coupled lines~\cite{gambetta:17}. The most powerful and flexible connectivity would be a complete graph, in which each qubit can couple with any other. A good compromise between these extremes would be a planar lattice, such as a square lattice, as required for the implementation of the most prominent error correcting codes \cite{dennis,lidar:13}.

\begin{figure}
    \centering
    \begin{subfigure}[b]{0.5\columnwidth}
        \includegraphics[width=\textwidth]{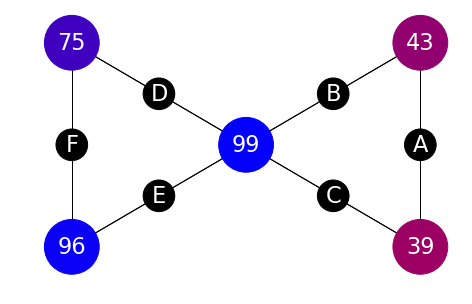}
        \caption{5 qubit IBM device $\mathtt{ibmqx4}$}
    \end{subfigure}
    \begin{subfigure}[b]{\columnwidth}
        \includegraphics[width=\textwidth]{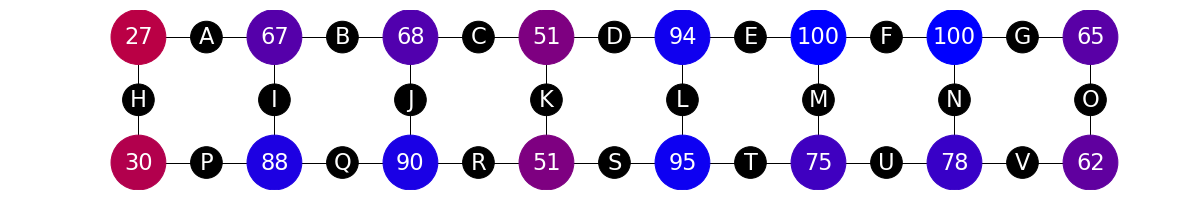}
        \caption{16 qubit IBM device $\mathtt{ibmqx5}$}
    \end{subfigure}
    \begin{subfigure}[b]{0.5\columnwidth}
       	\includegraphics[width=\textwidth]{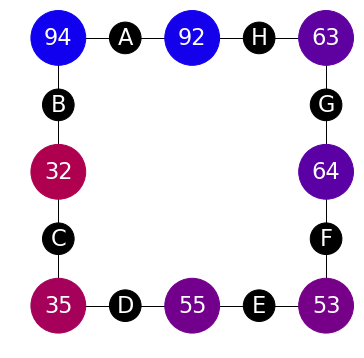}
       	\caption{8 qubit Rigetti device $\mathtt{8Q-Agave}$}
    \end{subfigure}
    \begin{subfigure}[b]{\columnwidth}
        \includegraphics[width=\textwidth]{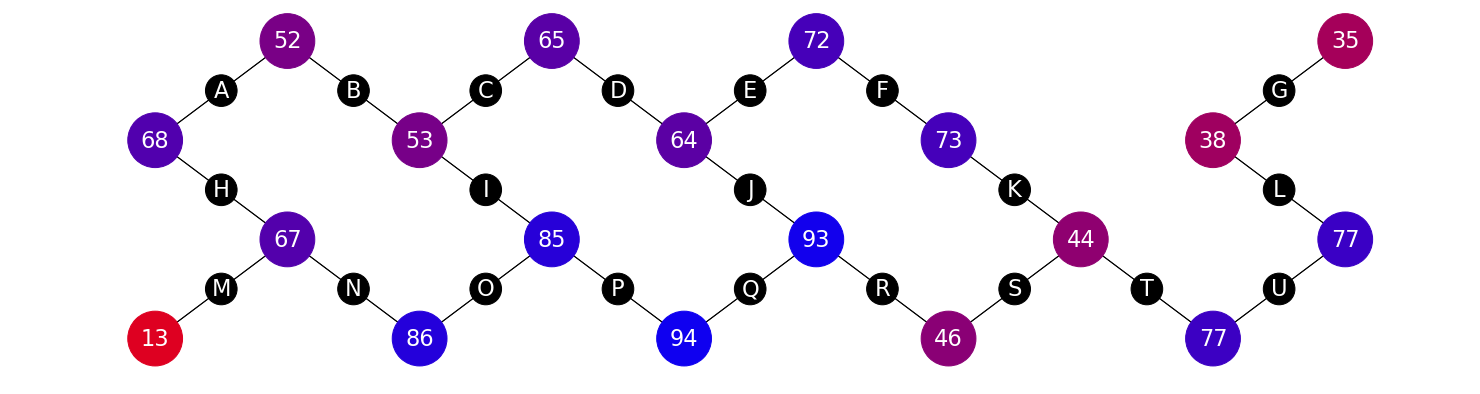}
        \caption{19 qubit Rigetti device $\mathtt{19Q-Acorn}$}
    \end{subfigure}
    \caption{Coupling graphs for devices studied in this work. Coloured circles denote qubits. The lines between them (labelled with letters) denote pairs for which entangling gates can be performed. The numbers within the coloured lines show an example set of results for the circuit, with each representing $\tilde{\theta_j} / (\pi/2)$ (as defined in Eq. (\ref{angle})) expressed as a percentage.}\label{fig:graphs}
\end{figure}

The competition between what can be easily implemented and what is required will lead to a range of different connectivity graphs being explored in near-term devices. As such, our benchmarking must be tailored to the specific capabilities of each device. The graphs for the devices used in this study are shown in Fig. \ref{fig:graphs}.

For a given connectivity graph, we will consider random circuits generated by sampling gates from the available gate set. The method we propose to do this is designed such that: (i) the rate at which entanglement builds up can be made as slow as desired, and (ii) the output possesses easily recognizable structure for states with only simple entangled states. This allows us to gain insights by comparing runs with different rates of entanglement generation, and use the loss of the structure in the output to assess how the entanglement build-up occurs.

To do this, we build up circuits as a series of \textit{rounds}. Each of these is composed of a pair of \textit{slices}, known as the entangling slice and the inverse slice. The former is a randomly generated set of gates that entangle disjoint pairs of qubits, while the latter is an attempt to invert this. The quality of these attempted inversions determine how fast the entanglement builds up. In the extreme that the inversions are perfect, the state will always maintain only short-range entanglement. In the extreme the inversions are generated without reference to the gates they are supposed to be inverting, they will help accelerate the build-up of long-range entanglement.

Typically, we consider the inversion gates to be chosen based on output data from an implementation of the circuit so far. Specifically, the first run of the circuit uses only the entangling slice of the first round. The results are then used infer the form of this entangling slice and propose an inverse. The next run then implements the whole of the first round, followed by the entangling slice of the second. The results are used to define the inverse slice of the second round, and so on.

The random generation of gates in each entangling slice is done by first randomly choosing a set of disjoint pairs of qubits. This pairing should be based on the connectivity of the device used: it should be possible to directly implement a controlled gate between the two qubits of each pair. The pairing is therefore a matching of the connectivity graph. In the case that the device has an odd number of qubits, or if the connectivity requires it, some qubits can be left unpaired.

Next, a random entangling gate is generated for each pair of qubits, $(j,k)$. This will take the form

\be
\mathtt{cx} (j,k) \, \exp [ \, i \, \theta_{jk} \, \sigma^j_x \, ] \, \mathtt{cx}(j,k) = \exp [ \, i \, \theta_{jk} \, \sigma^j_x \sigma^k_x \, ]
\ee

Here $\mathtt{cx}(j,k)$ denotes a controlled-NOT with $j$ as control and $k$ as target. The angles $\theta_{jk}$ are chosen randomly from the range $\pi/40 \leq \theta_{jk} \leq \pi/4$. The effect of these gates on an initial state with $\ket{0}$ for all qubits will be to create entangled pairs of the form
\be \label{state}
\cos (\theta_{jk}) \ket{00} + i \sin (\theta_{jk}) \ket{11} .
\ee
For such pairs, note that $Z$ basis measurement of the two qubits will always yield the same result. The probability that this result is $\ket{1}$ is
\be \label{eq:prob}
p_j = p_k = \sin^2 (\theta_{jk}).
\ee
Since the $\theta_{jk}$ are restricted to the range $\pi/40 \leq \theta_{jk} \leq \pi/4$, the values of $p_j$ will yield values of $p$ between $0.006$ and $0.5$. The lower bound was chosen to ensure a degree of distinction between qubits involved in pairs (and therefore an entangling gate) and those left unpaired (and therefore with no gate applied).

Given this structure in the output, it should be possible to deduce the random gates applied using only the values of $p_j$ for each qubit. Alternatively we could also invert Eq. (\ref{eq:prob}) to obtain the following set of values for each qubit.
\be \label{angle}
\theta_j = \sin^{-1}( \sqrt{p_j} ).
\ee
These can first be used to deduce the pairing using the fact that $p_j=p_k$, and therefore $\theta_j=\theta_k$, for each pair $(j,k)$. The values then directly give us the angle used in the corresponding entangling gate. Since this information completely specifies the gates of the entangling slice, it can be used to construct the corresponding inverse slice.

It is important to note the deduced inverse will not be a true inverse in general. Reasons for this include:
\begin{itemize}
\item Noise in the implementation, such as imperfect gates and decoherence, will perturb the measured $p_j$ from their ideal values;
\item The finite number of samples used to estimate the $p_j$ will lead to statistical noise;
\item Failures in the inverses of previous rounds will result in the entangling slice not being applied to the all $\ket{0}$ state, and so Eq. (\ref{eq:prob}) applies only approximately;
\item The use of a non-optimal to construct the inverses.
\end{itemize}


When the entangling slice of each round is not fully inverted, randomness will build up in the circuits. By choosing how strong these effects are, we can tune the rate at which long-range entanglement is generated.

Note that each round, as defined thus far, is completely diagonal in the $\sigma_x$ basis of all qubits. Using only such gates will not allow us to fully explore the Hilbert space of the device. The finishing touch for each round will therefore be to conjugate completed rounds with random single qubit gates. Each of these is randomly chosen to be either an $x$ or $y$ axis rotation, and with a randomly chosen angle $0\leq\phi\leq\pi/2$.

\section{Figures of merit}

With the results from running the circuit for each round we can assess the build-up of entanglement in a device. This will primarily be done by looking at how well the output can be used to deduce the inverse of the most recent slice of randomly chosen entangling gates. Highly successful construction of the inverse implies that long-range entanglement is negligible, and that the state immediately prior to the most recent slice was close to the all $\ket{0}$ state. The relation of Eq. (\ref{eq:prob}), and all conclusions derived from it, will then hold to good approximation.

On the other hand, highly unsuccessful construction of the inverse implies that final output is dominated by other effects. In the best case, this will be long-range entanglement built up by the random circuit. In the worst case, it will be noise. By comparison of random circuits for which entanglement is generated at different rates, we can attempt to distinguish these to possibilities.

Note that we will use $\tilde{p_j}$ to denote the measured probability of qubit $j$ to output the result $\ket{1}$. As the \textit{measured} value, this is distinct from the true value $p_j$ in general, because of the effects of the imperfections listed in the previous section.

\subsection{Fuzz}

The first way we will quantify an output is to compare the calculated values of $\tilde{p_j}$ and $\tilde{p_k}$ for each pair in the most recent slice. If Eq. (\ref{eq:prob}) holds, we will have $\tilde{p_j}=\tilde{p_k}$ in each case. However, as Eq. (\ref{eq:prob}) becomes an increasingly worse approximation, these numbers will begin to drift away from each other. We refer to this as \textit{fuzz}, and quantify it as follows over the whole device
\be
\mathtt{fuzz} = \frac{ \sum_{(j,k)} | \tilde{p_j} - \tilde{p_k} | }{ n }.
\ee
Here $n$ is the number of pairs of qubits on the device (and so half the total number of qubits when this is even and the connectivity allows).

Note that, for the first round, the fuzz will be at or close to zero. It will then begin to rise as Eq. (\ref{eq:prob}) becomes more approximate. At the other extreme, after an arbitrarily large number of rounds, all $\tilde{p_j}$ will converge at close to $1/2$. This will ideally be due to the random circuit causing the final state to be a typical sample drawn uniformly from the multiqubit Hilbert space. However, it could also be due to the build up of noise. In either case, the fact that all $\tilde{p_j}$ have converged to the same value will also cause a low value of the fuzz.

Given this behaviour, a graph of fuzz against round number will necessarily feature a peak. This will be the most noticeable feature in our results. It essentially marks the start of the inevitable march towards a completely random output without the structure required for inverses to be successfully deduced.

We will look at the build-up of fuzz for each device in two different cases:
\begin{enumerate}
\item Inverses constructed when the assumed pairing of the qubits is completely correct, and the deduced angles are correct up to effects caused by statistical noise and the build-up of entanglement;
\item Inverses constructed when the assumed pairing is chosen completely randomly and without reference to the results.
\end{enumerate}

For simulated instances of case 1, and for case 2, the assumed $\theta_{jk}$ are calculated from Eq. (\ref{eq:prob}) using $(\tilde{p_j}+\tilde{p_k})/2$. This not done when we consider case 1 to be run on a real device, since the presence of noise would lead to low quality inverse slices. The effects of statistical noise alone is therefore emulated by taking the correct values and adding $0.1/\sqrt{\mathtt{shots}}$, where $\mathtt{shots}$ is the number of samples used for $\tilde{p_j}$. 

\begin{figure}[ht]
	\centering
    \includegraphics[width=\columnwidth]{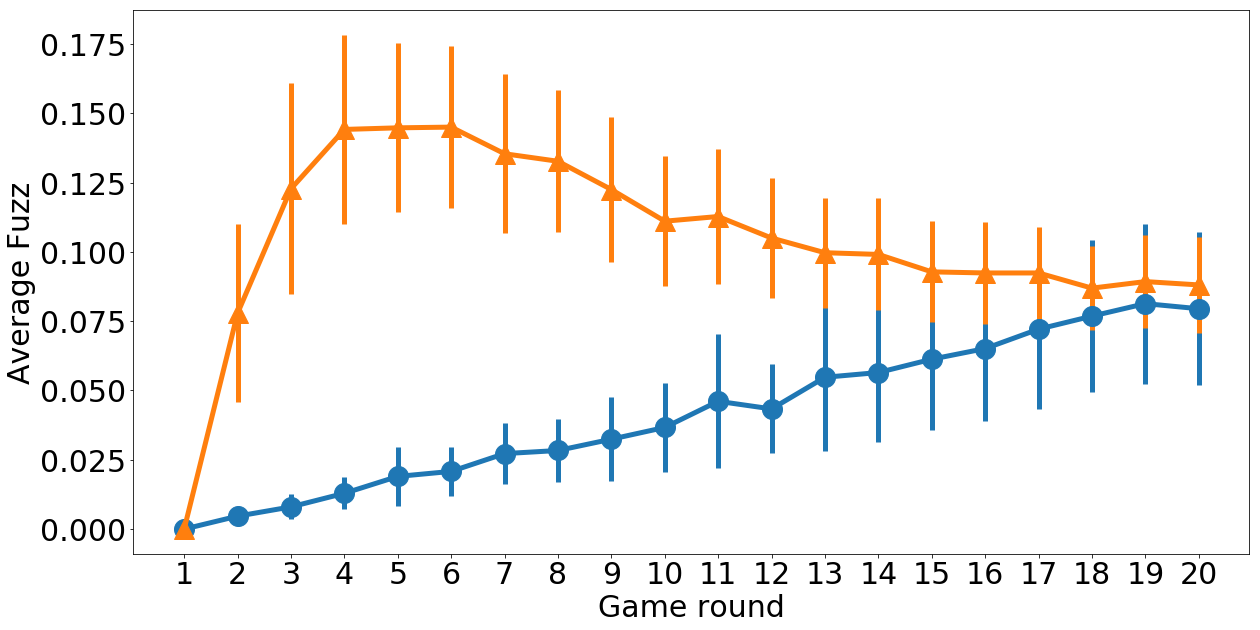}
	\caption{Examples of how fuzz builds up over the rounds for (orange) pairs in the inverse slices that are chosen randomly and (blue) those chosen correctly with $\mathtt{shots}=100$.}\label{fig:fuzz}
\end{figure}

In the absence of noise, the inverses for case 1 are perfect up to statistical noise. This can be suppressed arbitrarily by increasing the number of samples used to calculate the $\tilde{p_j}$. For case 2, however, the fuzz will rise sharply and peak early. This is because the second slice of each round is essentially as much a source of random gates as the first, and has little effect as an inverse. Examples of these graphs are shown in Fig. \ref{fig:fuzz}.

Building up random circuits able to uniformly sample states from the multiqubit Hilbert space requires the fuzz to first peak, and then subsequently vanish. However, this same behaviour will also be seen for devices dominated by noise. It is therefore important to determine which of these two possibilities occurred when such a peak is observed. To do this, we can run the process again over the same number of rounds, but instead use an instance of the random circuits for which entanglement builds up much more slowly. So if we are considering the peak resulting from case 2 (inverse slices with randomly chosen pairs), we can additionally study case 1 (inverse slices with correctly chosen pairs). If the noise is dominant, the fuzz will again be seen to peak and vanish. If noise is negligible, however, and a large value of $\mathtt{shots}$ is used, the increase in fuzz will be much slower.

Ideally, we would like to see the fuzz remain at a low and pre-peak value for case 1 for as many rounds as it takes for the fuzz of case 2 to first peak and then subsequently vanish. Satisfying this condition would then provide strong evidence that the vanishing fuzz for case 2 is primarily due to the build up of entanglement and not noise.


\subsection{Success rate for pairing}

We will now consider how well the pairing can be deduced for a given output. We will do this using minimum weight perfect matching (MWPM)~\cite{edmonds:65,rantwijk} on the connectivity graph of the device. For each pair of qubits we assign a weight that depends on the values of the $\tilde{\theta_j}$ derived via Eq. \ref{angle} from the measurement values of the $\tilde{p_j}$,
\be \label{weight}
W_{j,k} = | \tilde{\theta_j} - \tilde{\theta_k} | .
\ee
The minimum weight matching will find the pairing that minimizes this weight, and therefore minimizes the differences between the $\tilde{\theta_j}$ values. Since the $\tilde{p_j}$ values for the two qubits within each pair should be equal, performing this minimization should provide a near optimal means of finding the pairs. The fraction of pairs correctly found my this method will be used as a further way of analysing the progress of our random circuits.

\begin{figure}[ht]
	\centering
    \includegraphics[width=\columnwidth]{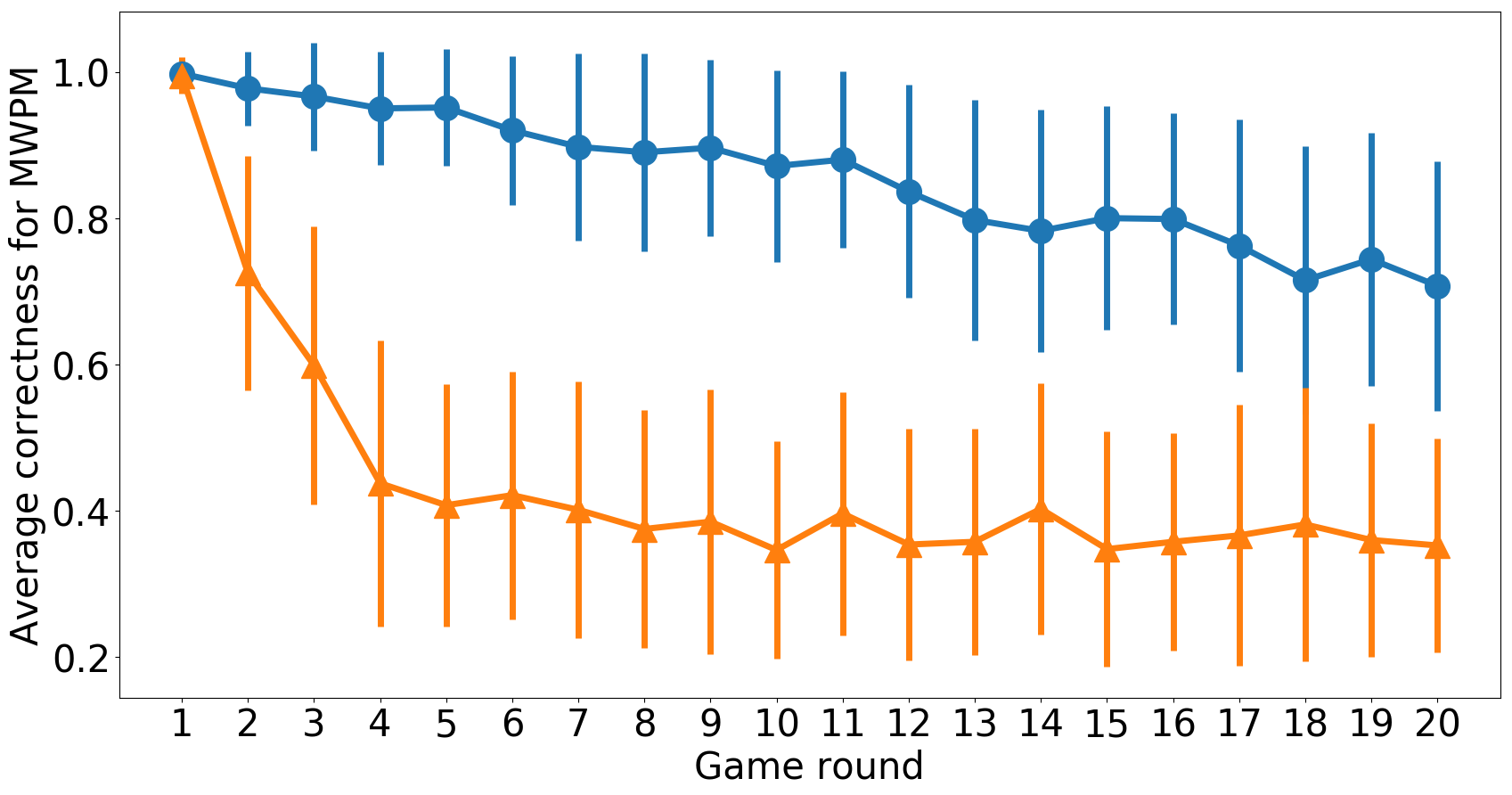}
	\caption{Examples of how the success rate for MWPM decays over the rounds for (orange) pairs in the inverse slices that are chosen randomly and (blue) those chosen correctly with $\mathtt{shots}=100$.}\label{fig:mwpm}
\end{figure}

Examples of how this success rate will decay over the rounds are shown in Fig. \ref{fig:mwpm}.

Note that the success rate for MWPM should ideally be $100\%$ for round 1, since the initial state is all $\ket{0}$ and Eq. (\ref{eq:prob}) will hold exactly. This will occur for inverses chosen by any method, because no inverses have yet been applied in this round. This initial value of the success rate will therefore be a particularly interesting point to consider.

\subsection{Difference with ideal values}

The main result taken from the output is the set of measured probabilities $\tilde{p_j}$. It therefore makes sense to compare these values directly to the ideal values $p_j$. Specifically, we will calculate the difference between the corresponding $\tilde{\theta_j}$ and the actual angle $\theta_{jk}$ used for the pair that qubit $j$ is a part of. This will be averaged over the entire device to give a measure of how well the outputs of the qubits correspond to what would be expected from the most recent entangling slice alone.

\be
\mathtt{diff} = \frac{ \sum_{(j,k)} | \tilde{\theta_j} - \theta_{jk} | + | \tilde{\theta_j} - \theta_{jk} | }{ 2 n }.
\ee

\begin{figure}[ht]
	\centering
    \includegraphics[width=\columnwidth]{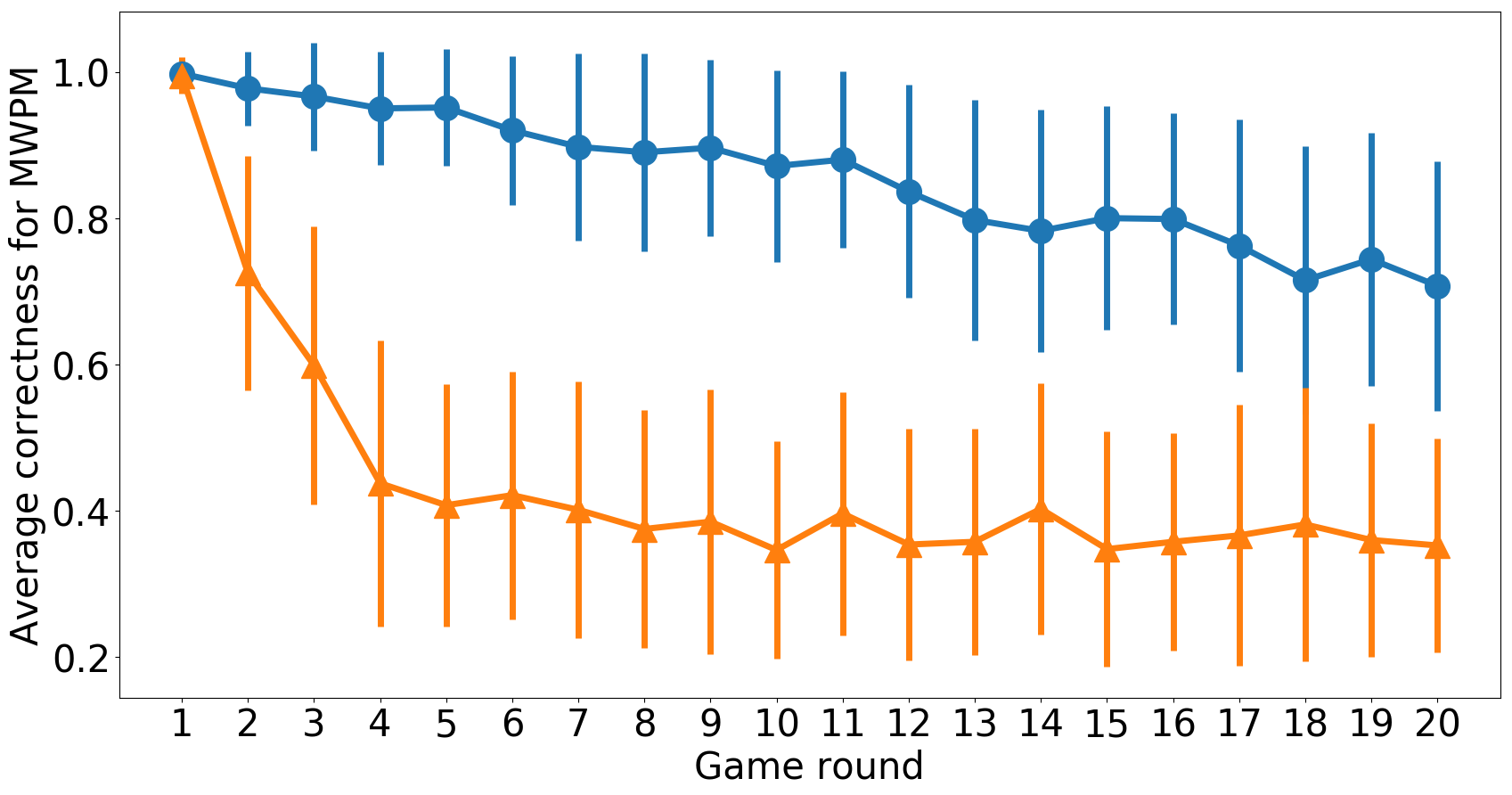}
	\caption{Examples of how the difference between the inferred and actual  $\theta_{jk}$ builds up over the rounds for (orange) pairs in the inverse slices that are chosen randomly and (blue) those chosen correctly with $\mathtt{shots}=100$.}\label{fig:diff}
\end{figure}

Examples of how this difference will build up over the rounds are shown in Fig. \ref{fig:diff}.

\subsection{Error mitigation}

Thus far we have used only one of the properties of the states described in Eq. (\ref{state}): the fact that $p_j=p_k$. However, it is further true that the results for paired qubits $j$ and $k$ should be perfectly correlated. This provides a further means by which pairs can be identified from the data. Specifically, the mutual information $I(j;k)$ for measurement outcomes will be non-zero if and only if the qubits are paired. For a given qubit $j$, the qubit $k′$ with which we can expect it to be paired is therefore that with the highest value of $I(j;k′)$. Let us refer to this as qubit $c(j)$.

This information could then be used mitigate for effects that cause violations of Eq. (\ref{eq:prob}). Specifically, instead of using the measured values of the $\tilde{p_j}$, we can instead use
\be
\bar{p_j} = \frac{ \tilde{p_j} + \tilde{p_{c(j)}} }{2}
\ee
For cases in which two qubits are each most correlated with the other (i.e. $c(j)=k$ and $c(k)=j$), the resulting values of $\bar{p_j}$ and $\bar{p_k}$ will be equal, and so satisfy one of the expected behaviours from Eq. (\ref{eq:prob}). Assuming that the mutual informations can be used to correctly deduce pairings in most cases, this will result in significant improvements to results.

\subsection{Quantum Awesomeness}

Determining the most likely pairing of the qubits given the $\tilde{p_j}$ (or error mitigated $\bar{p_j}$) is a puzzle to be solved. Indeed, it is a puzzle that can be played even without knowledge of the underlying quantum programming. Our scheme then becomes an accessible puzzle game.

If played directly on a device (real or simulated) the pairing supplied by a player will be used to construct the inverse slice for each round. This means that any mistakes made will have an effect on all subsequent rounds. This, as well as other effects which cause the build-up of long-range entanglement or noise, will cause the puzzles to increase in difficulty for each successive round. The aim of the player will then be simply to keep the game playable for as long as possible. Quantum computational supremacy with random circuits then corresponds to allowing a player to reach a unplayably hard \textit{Game Over} state that is caused by the onset of the Porter-Thomas regime.

The game can also be played using saved data, such as that from a run in which the pairs of the inverse slice are always chosen correctly. In this case, the main purpose of the game is to serve as a qualitative way of benchmarking devices. Greater size and more complex connectivity will allow more challenging puzzles, whereas greater noise will cause an infuriating degree of difficulty. The quality of a given device will therefore correlate well to how enjoyable the game is when played on it. This allows a high-level means of comparing devices that is accessible by the interested lay person.

Examples of what is game would look like for the devices considered in this work are shown in Fig. \ref{fig:graphs}. In these, the number shown on each qubit is the corresponding $\tilde{\theta_j} / (\pi/2)$ expressed as a percentage. These numbers also determine the colour used for each qubit, ranging from blue for $0\%$ to red for $100\%$. The aim is therefore to identify the correct pairs (which are labelled by letters) by matching qubits with similar numbers and colours.

This game, which is called \textit{Quantum Awesomeness}, can be played with the data presented in this paper at~\cite{awesomeness}.

\section{Results}

Results were taken for a selection of real and example devices, with both real and simulated data. For runs on real devices, data is taken only for the case of correct pairings with a large number of shots ($\mathtt{shots}\sim 10000$). This is then compared to simulated data for the case with random pairings, and for that of correct pairings but far fewer shots ($\mathtt{shots} = 100$). Ideally, the results should show a build-up of entanglement that is slower than for both the simulated instances. It should especially be much less than for the simulated case of random pairing.

To provide a good understanding of how the figures of merit should behave, we first consider simulated results from a set of example devices.

\subsection{Example devices} \label{subsec:example}

The quantitative benchmarks of the previous section were applied to a set of example devices of different sizes and connectivities. The connectivity graphs considered were a line (with 5, 11, 15 and 19 qubits), a ladder (4, 10, 16 and 20 qubits), a square lattice (4, 9 and 16 qubits) and a fully connected graph (5, 11, 16 and 19 qubits). The qubit numbers where chosen to span the same range as the real devices we will consider, given the constraints of the connectivity (square numbers only for the square lattice, etc).

\begin{figure*}
    \centering
    \begin{subfigure}[b]{\textwidth}
        \includegraphics[width=0.9\textwidth]{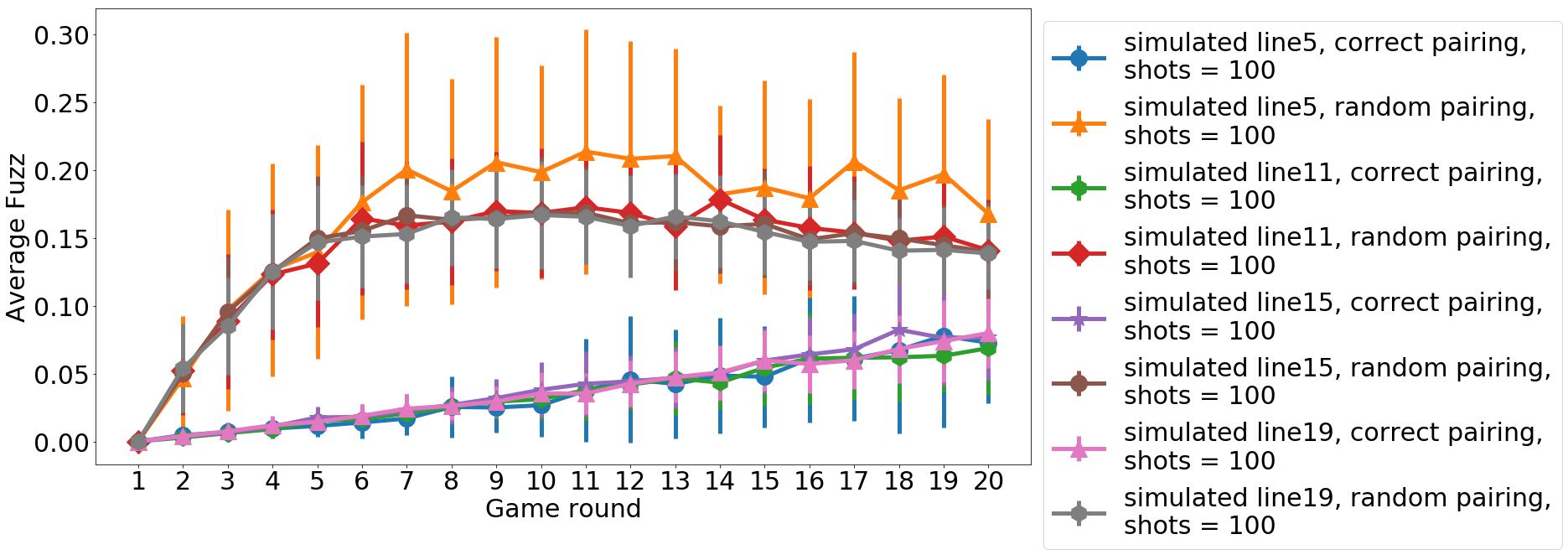}
    \end{subfigure}
    \begin{subfigure}[b]{\textwidth}
        \includegraphics[width=0.9\textwidth]{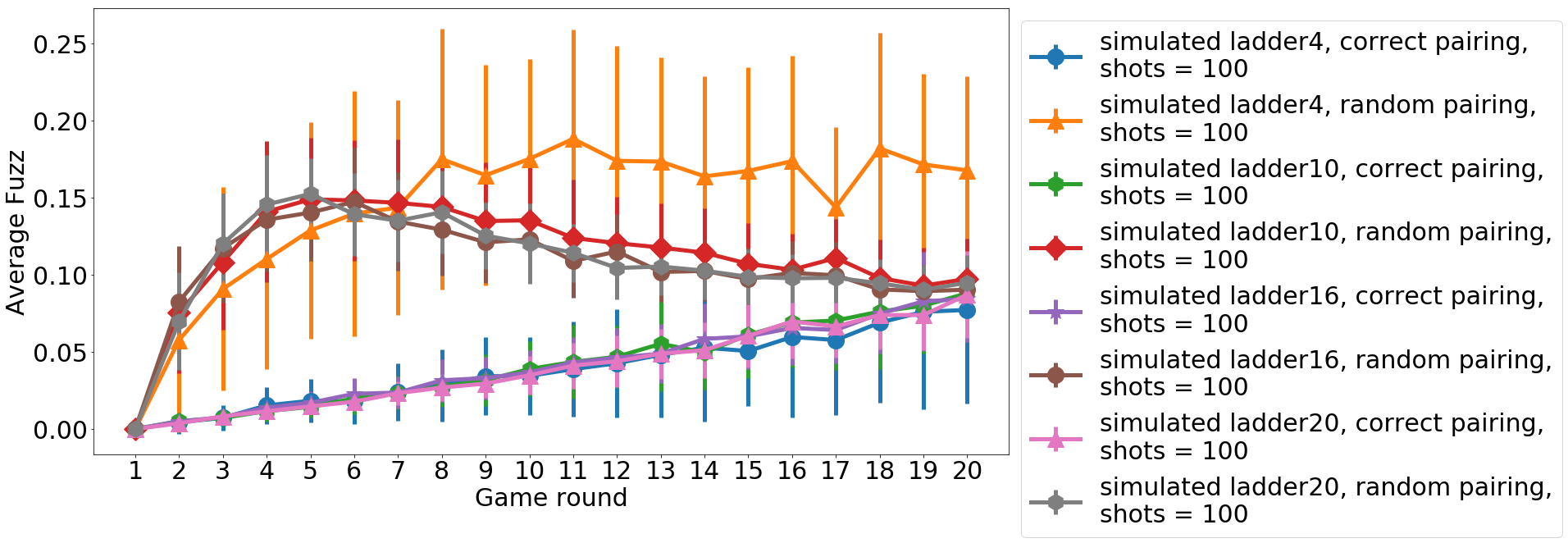}
    \end{subfigure}
    \begin{subfigure}[b]{\textwidth}
        \includegraphics[width=0.9\textwidth]{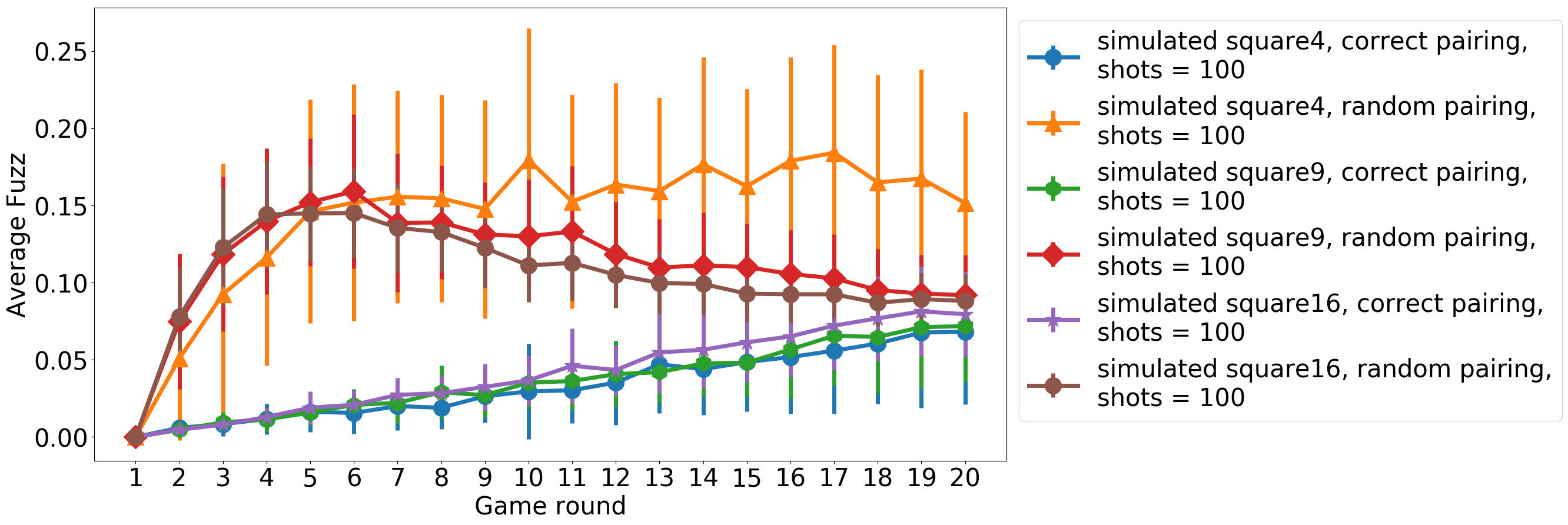}
    \end{subfigure}
    \begin{subfigure}[b]{\textwidth}
        \includegraphics[width=0.9\textwidth]{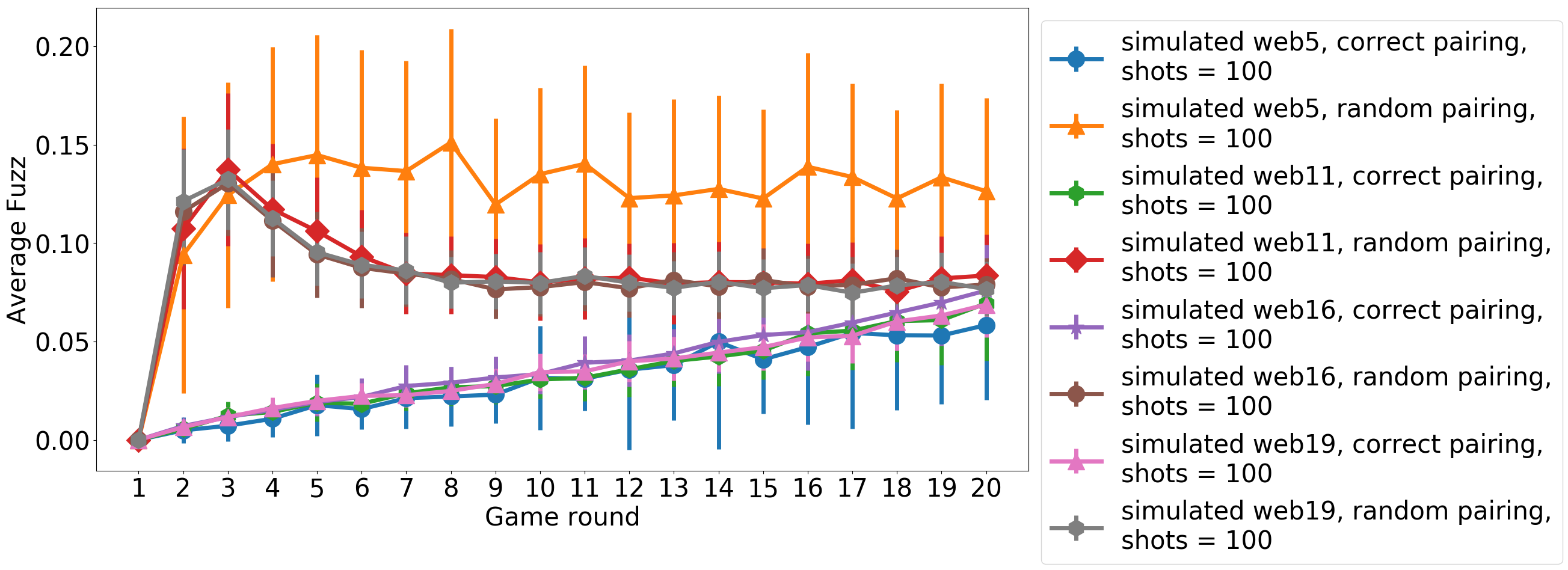}
    \end{subfigure}
    \caption{The average fuzz for all example devices. Each point is the average of 100 samples, with error bars given by the standard deviation. These results are discussed in section.}\label{fig:example_fuzz}
\end{figure*}

Results for the fuzz are shown in Fig. \ref{fig:example_fuzz}. For each connectivity, the fuzz for the smallest case was found to behave very differently than the others. This is due to the fact that the fuzz will converge to a value that decays exponentially with qubit number as the state fully explores the Hilbert space. So though it can be said to vanish for large devices, it will not for small ones. The peak for the fuzz is therefore less visible for such small sizes.

For the larger devices, the graphs show little variation for different sizes over the range considered. This is despite the fact that the build-up of long-range entanglement will scale with system size~\cite{boixo:18}. This shows that our figures of merit document the process of the entanglement moving beyond the simple pairing provided by the entangling slices, rather than it becoming truly long-range. The vanishing of the fuzz, for example, is therefore not a witness of the build-up of long-range entanglement, but is a necessary condition.

The peak and subsequent decay of the fuzz is found to depend strongly on connectivity. These processes are slowest for the least connected devices (the lines) and fastest for the most connected devices (fully connected). The ladders and square lattices show similar behaviour, which lies between the two extremes.

\begin{figure*}
    \centering
    \begin{subfigure}[b]{\textwidth}
        \includegraphics[width=0.9\textwidth]{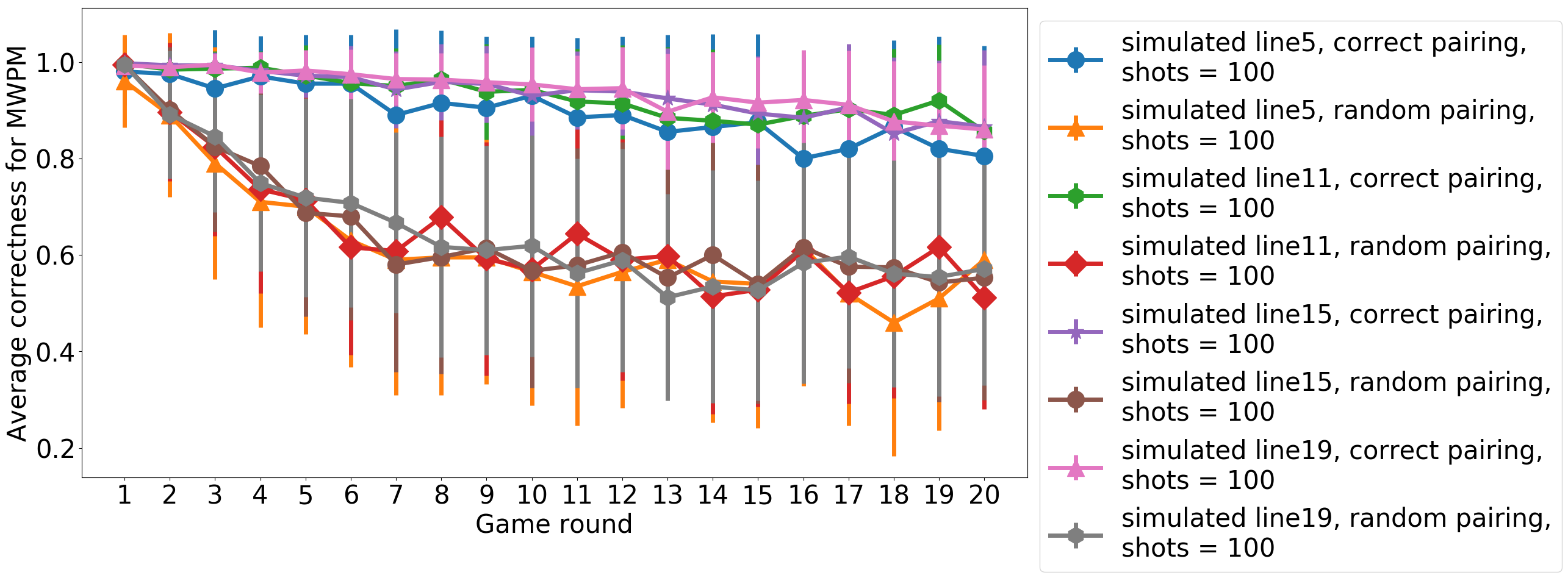}
    \end{subfigure}
    \begin{subfigure}[b]{\textwidth}
        \includegraphics[width=0.9\textwidth]{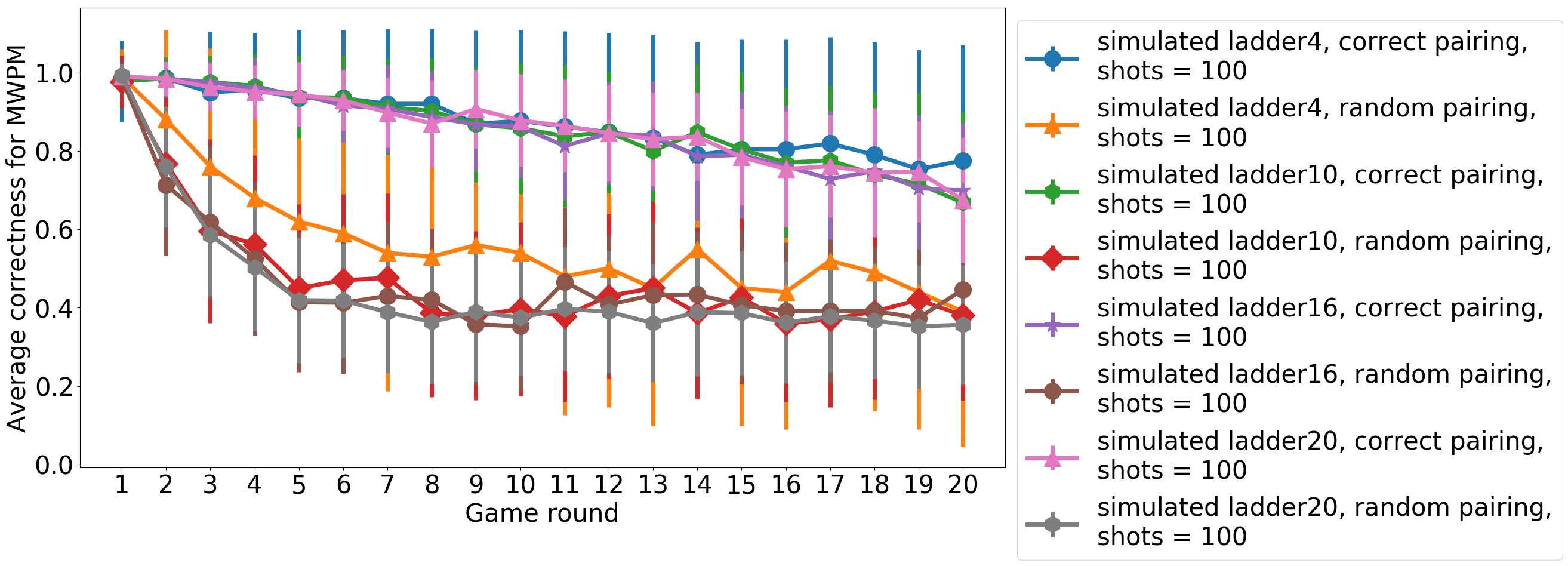}
    \end{subfigure}
    \begin{subfigure}[b]{\textwidth}
        \includegraphics[width=0.9\textwidth]{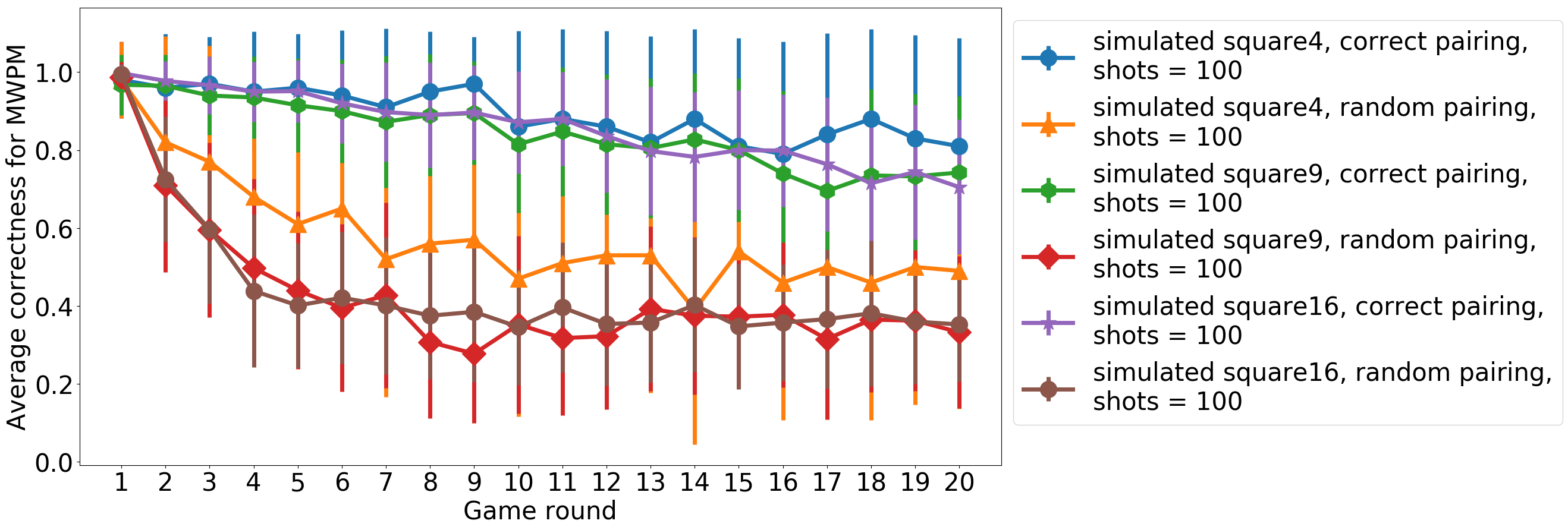}
    \end{subfigure}
    \begin{subfigure}[b]{\textwidth}
        \includegraphics[width=0.9\textwidth]{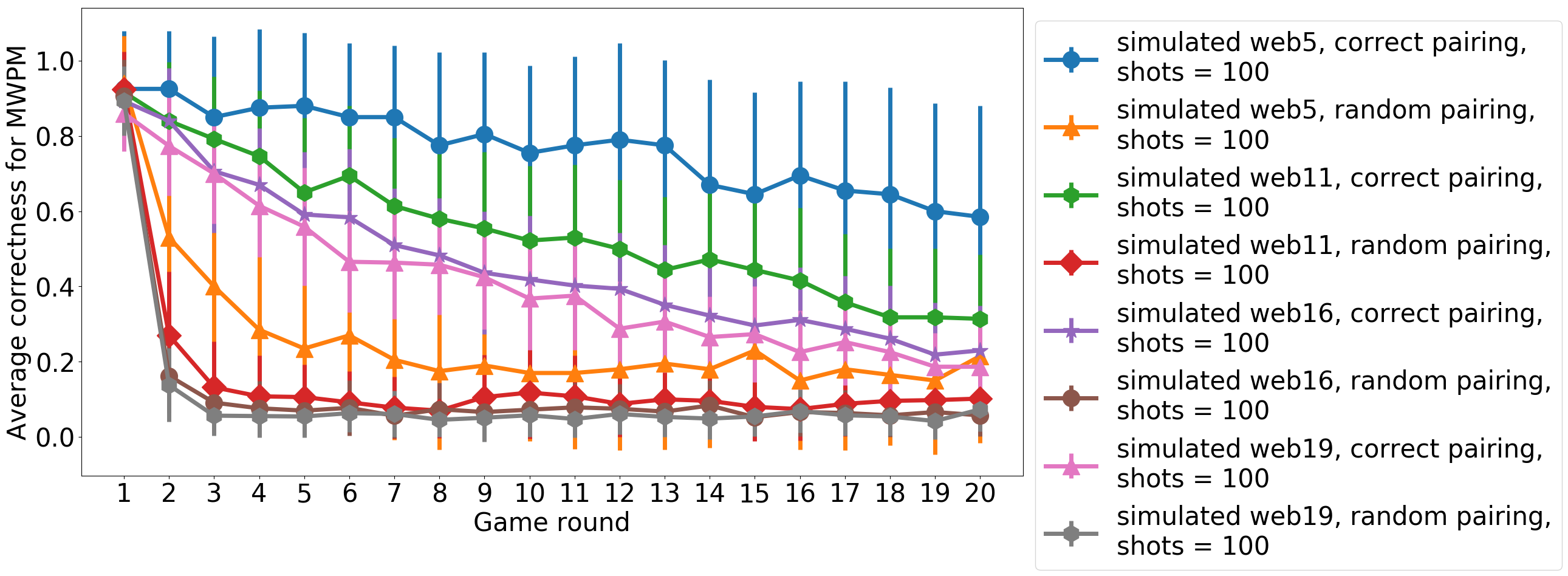}
    \end{subfigure}
    \caption{The average correctness of pairing via minimum weight perfect matching for all example devices. Each point is the average of 100 samples, with error bars given by the standard deviation.}\label{fig:example_mwpm}
\end{figure*}

Results for the success rate of MWPM are shown in Fig. \ref{fig:example_mwpm}. In each case, this fraction is found to decrease sharply for the first few rounds, before converging at a value which reflects the fraction of pairs that would be correct for a random guess. The round at which this occurs appears to correspond well to that at which the fuzz peaks. This further shows that this is a significant point at which the entanglement moves beyond the short-range correlations created directly by single entangling slices.

\begin{figure*}
    \centering
    \begin{subfigure}[b]{\textwidth}
        \includegraphics[width=0.9\textwidth]{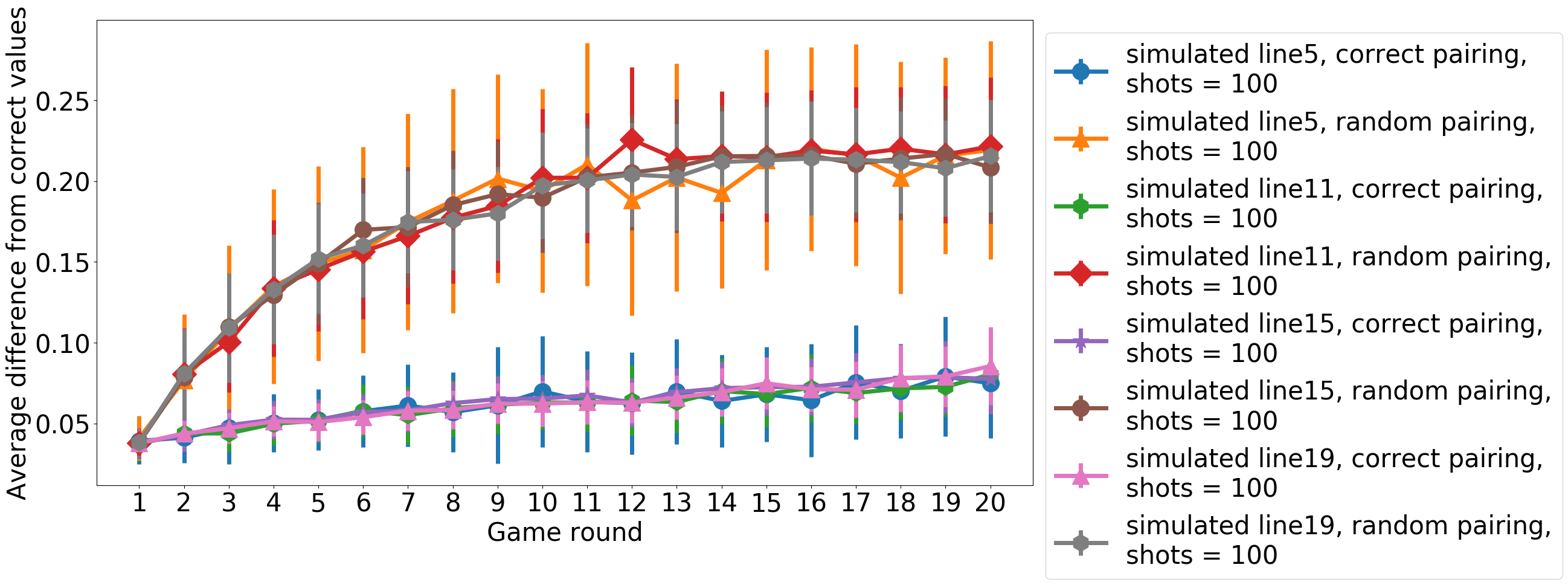}
    \end{subfigure}
    \begin{subfigure}[b]{\textwidth}
        \includegraphics[width=0.9\textwidth]{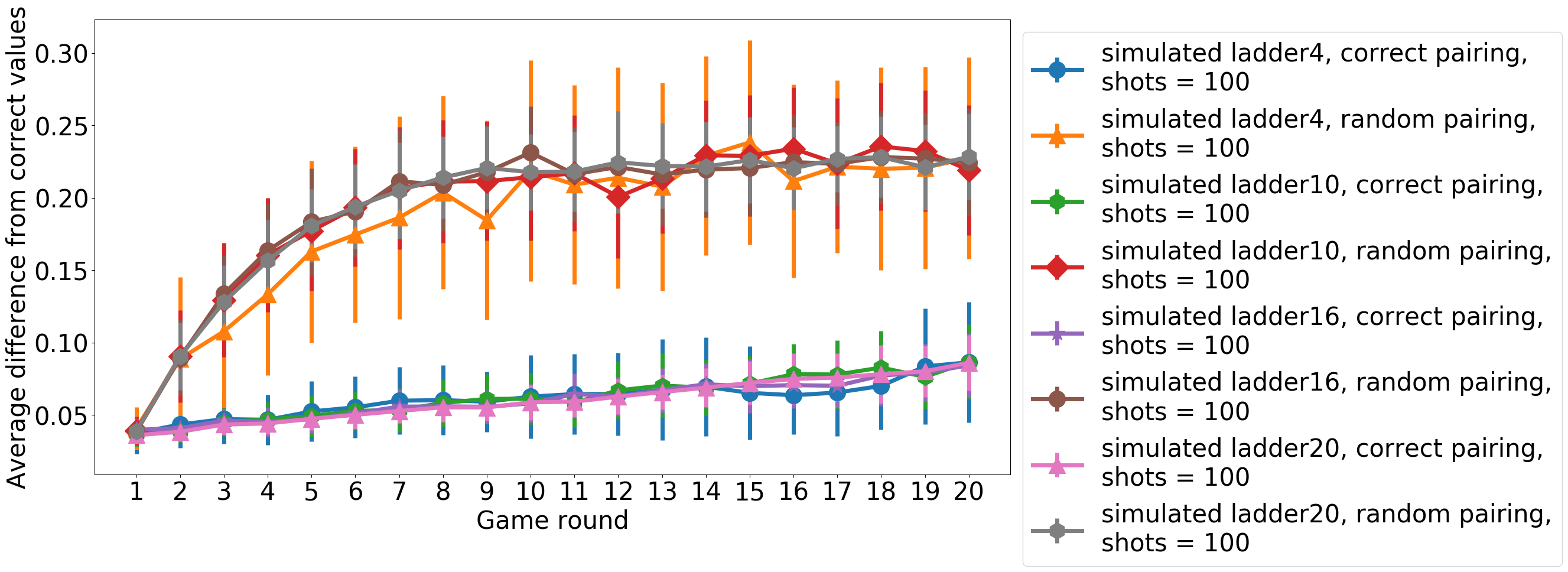}
    \end{subfigure}
    \begin{subfigure}[b]{\textwidth}
        \includegraphics[width=0.9\textwidth]{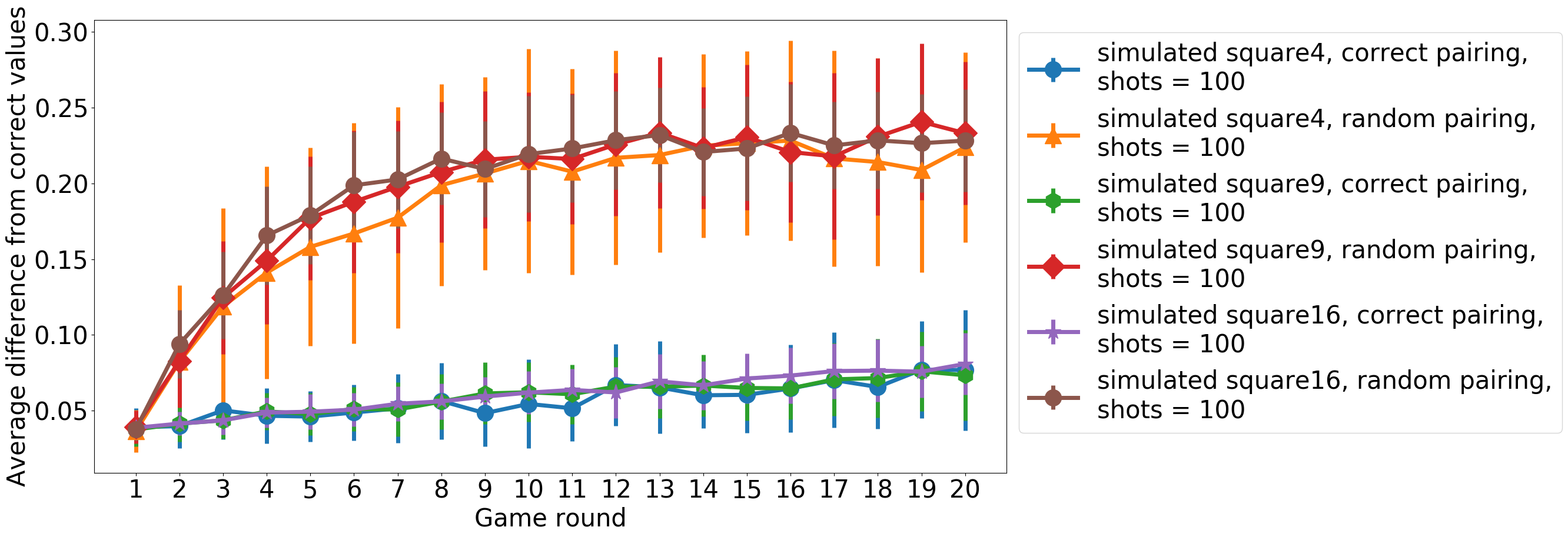}
    \end{subfigure}
    \begin{subfigure}[b]{\textwidth}
        \includegraphics[width=0.9\textwidth]{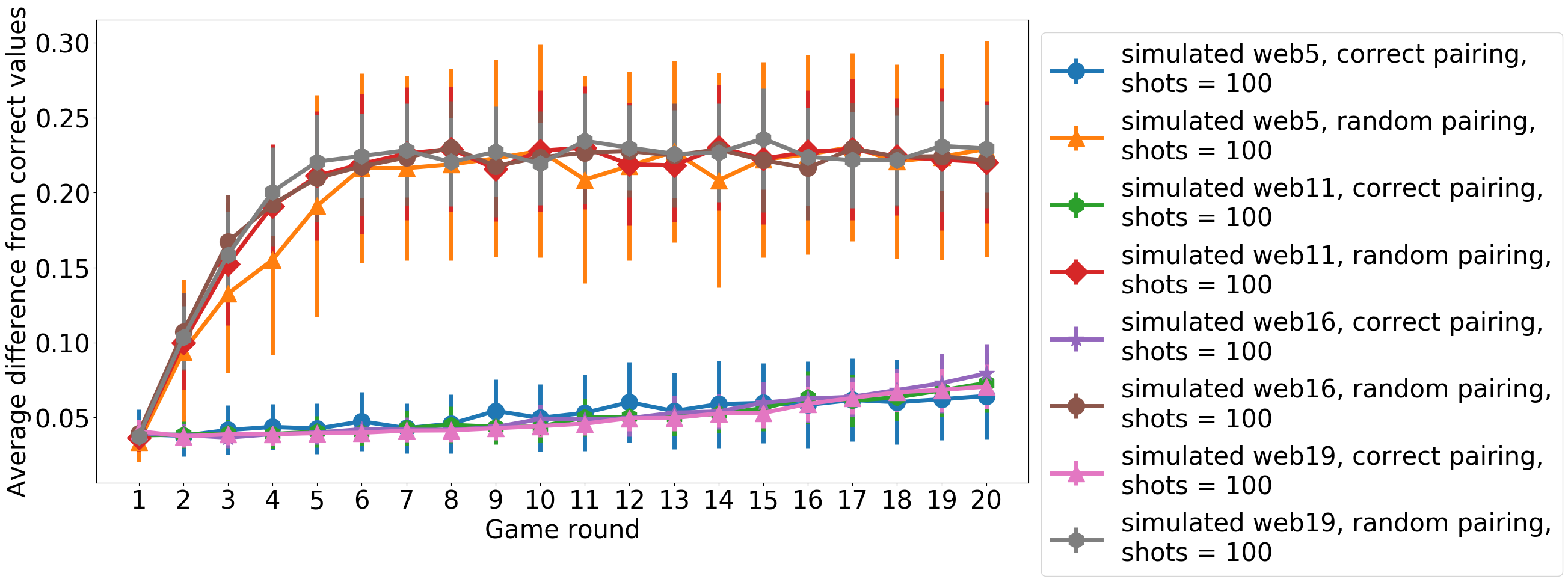}
    \end{subfigure}
    \caption{The average difference between inferred and correct values for the $\theta_{jk}$ for all example devices. Each point is the average of 100 samples, with error bars given by the standard deviation.}\label{fig:example_diff}
\end{figure*}

Similar behaviour is found for the difference between the $\tilde{\theta_j}$ and their ideal values, as shown in Fig. \ref{fig:example_diff}. It first rises sharply before showing signs of convergence. The peak of the fuzz is again found to be a good rule of thumb for the point at which this occurs.

\subsection{5 and 16 qubit IBM devices}  \label{subsec:ibm}

\begin{figure*}
    \centering
    \begin{subfigure}[b]{\textwidth}
        \includegraphics[width=\textwidth]{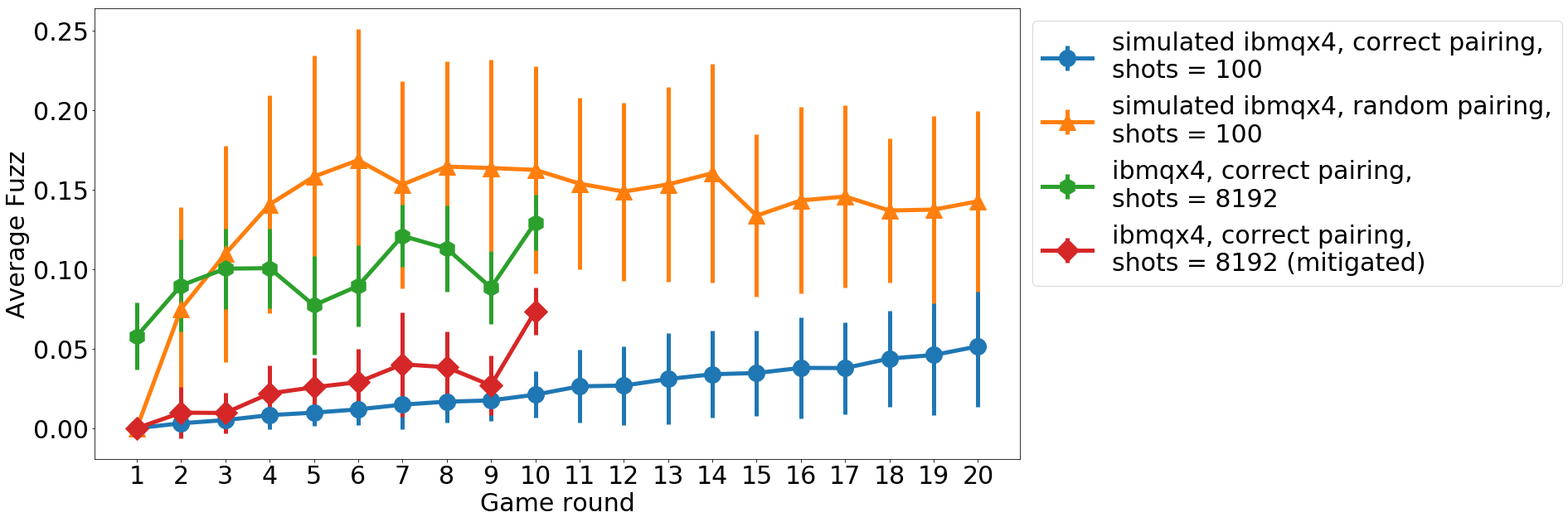}
    \end{subfigure}
    \begin{subfigure}[b]{\textwidth}
        \includegraphics[width=\textwidth]{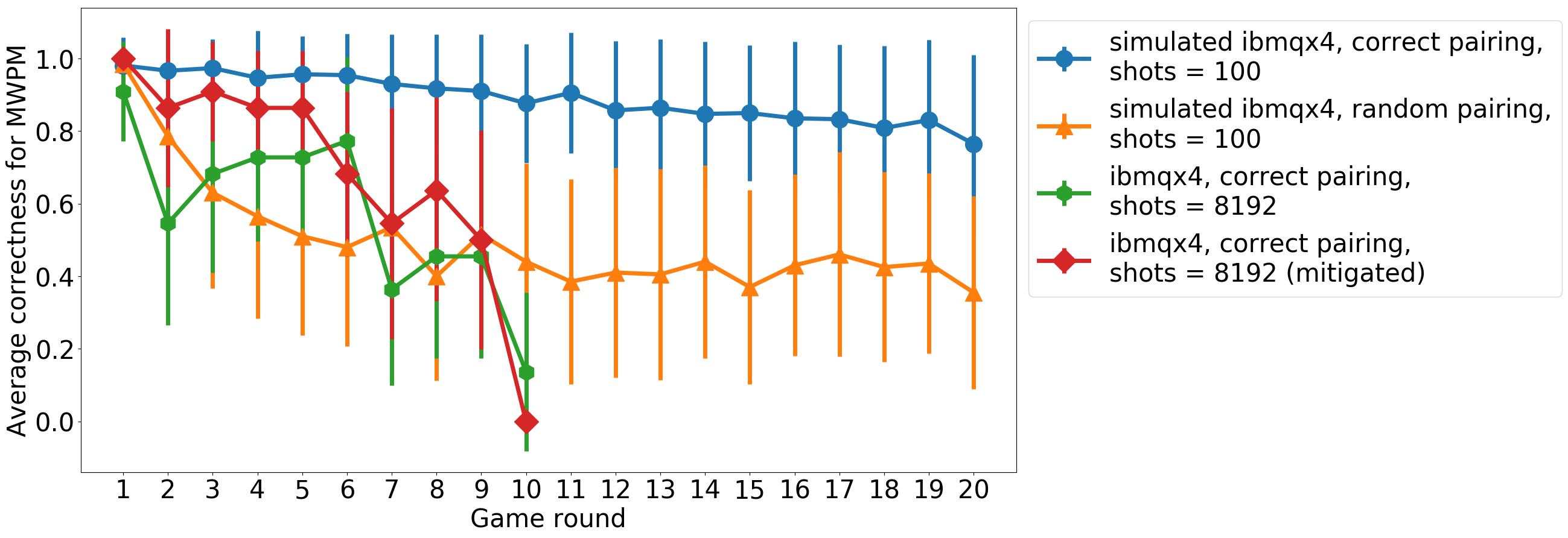}
    \end{subfigure}
    \begin{subfigure}[b]{\textwidth}
        \includegraphics[width=\textwidth]{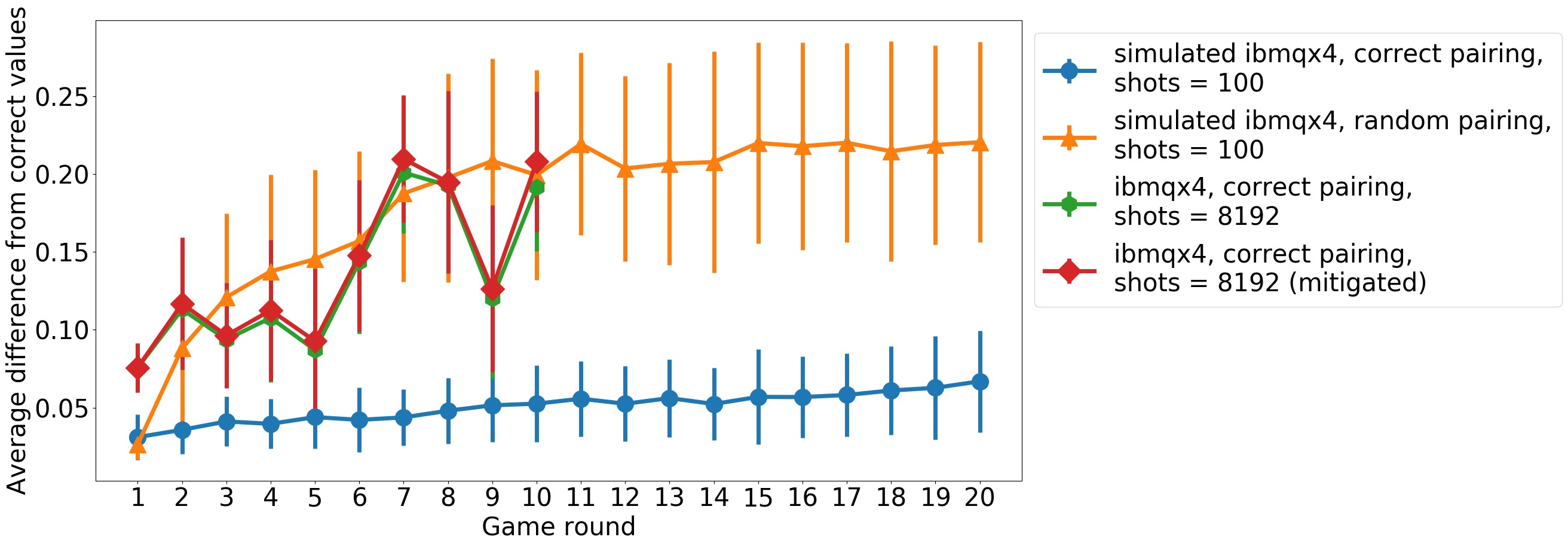}
    \end{subfigure}
    \caption{Results for the IBM device $\mathtt{ibmqx4}$. Each point is the average of 100 samples for simulated data. Error bars given by the standard deviation.}\label{fig:ibmqx4}
\end{figure*}

Results for the the 5 qubit IBM device $\mathtt{ibmqx4}$ are shown in Fig. \ref{fig:ibmqx4}. The small size of the device, and associated finite size effects, make it difficult to identify features such as the fuzz peak. 

One distinct feature, however, is that the fuzz for error mitigated results corresponds well to the simulated results for correctly chosen inverse slices up until around round 8.

For the success rate for MWPM, the round 1 value is reaches the expected $100\%$ when error mitigation is used, and also also remains high (over $80\%$) for the first few rounds. The success rate with non-mitigated data also exceeds $90\%$.

These results show that the expected structure in the output is well maintained for the first few rounds. Note that each round requires at least a depth of two controlled gates, and so the results show that the device maintains good coherence for even for circuits with a CNOT depth of 5 to 10.

Similar agreement is not seen for the average difference between the $\tilde{\theta_j}$ and their ideal values. However, for most round this is nevertheless found to be smaller for data from the real device than for the simulated case of  randomly chosen inverses. No improvement for this is achieved by the error mitigation, but this is because the method of mitigation used does not correct for these values.

\begin{figure*}
    \centering
    \begin{subfigure}[b]{\textwidth}
        \includegraphics[width=\textwidth]{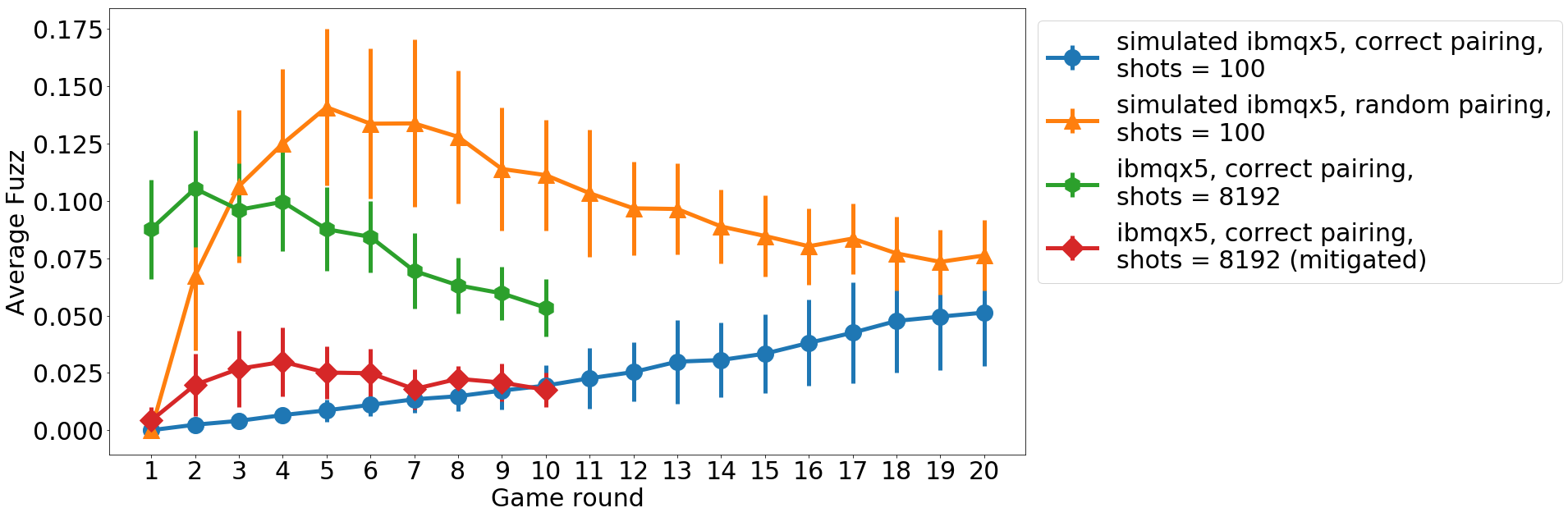}
    \end{subfigure}
    \begin{subfigure}[b]{\textwidth}
        \includegraphics[width=\textwidth]{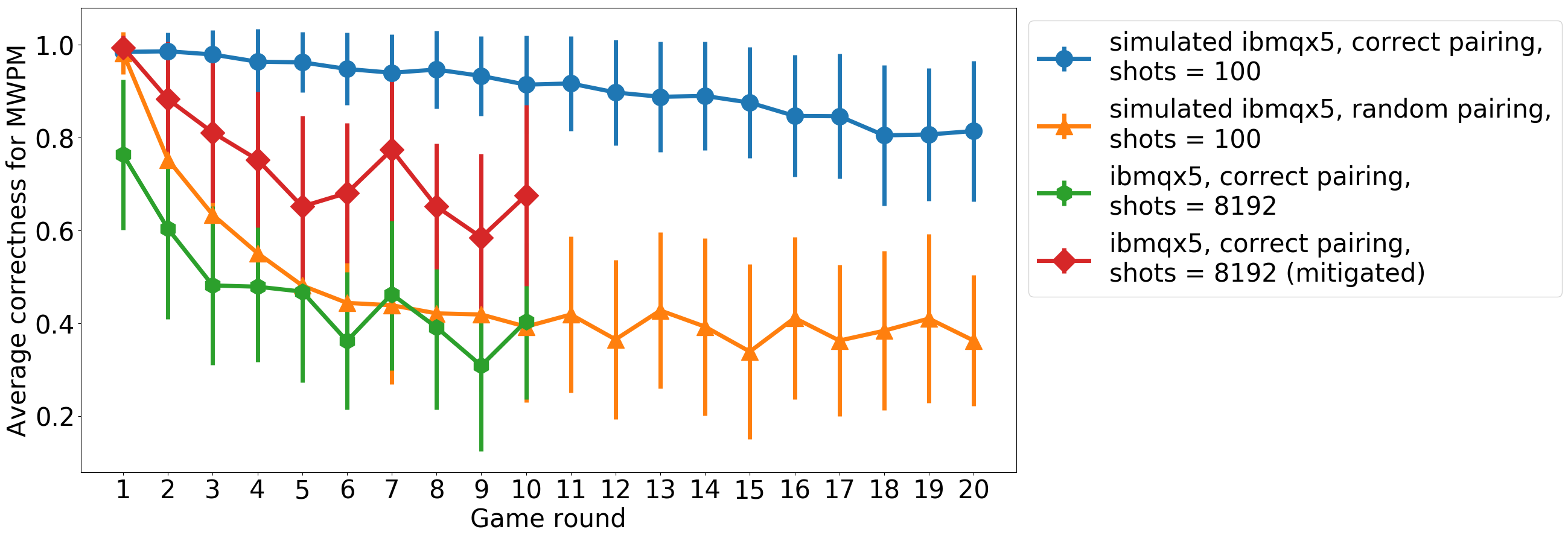}
    \end{subfigure}
    \begin{subfigure}[b]{\textwidth}
        \includegraphics[width=\textwidth]{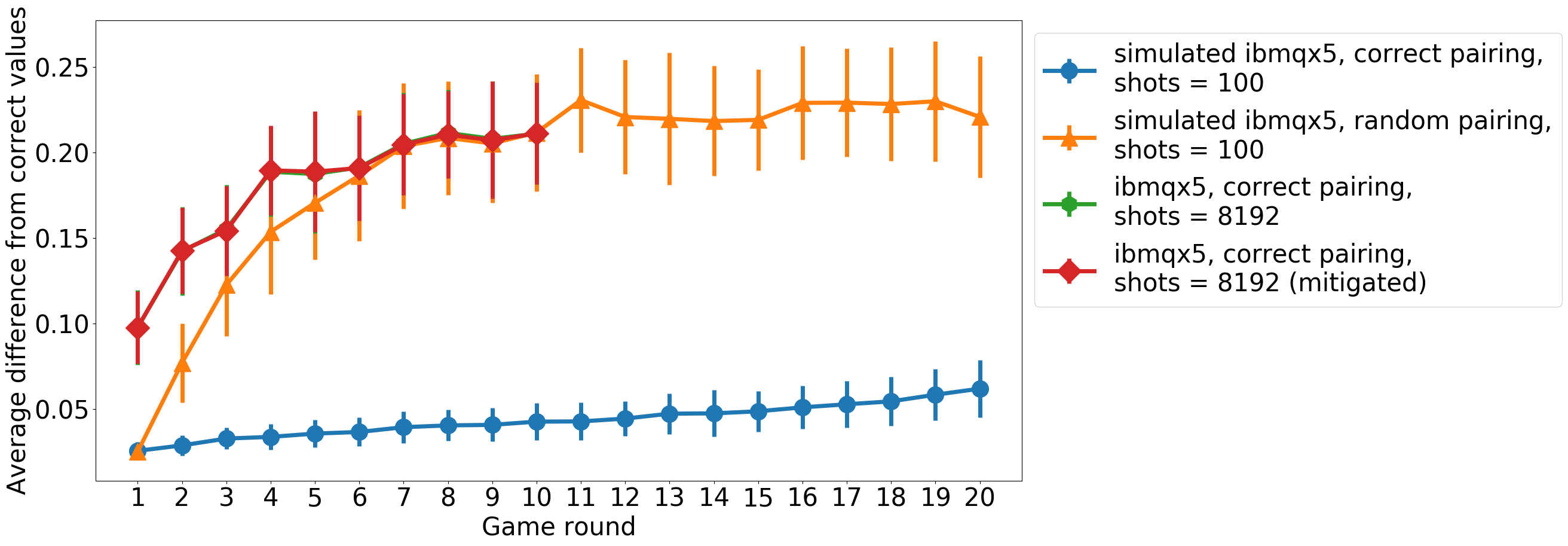}
    \end{subfigure}
    \caption{Results for the IBM device $\mathtt{ibmqx5}$. Each point is the average of 100 samples for simulated data and around 50 samples for the real device. Error bars given by the standard deviation.}\label{fig:ibmqx5}
\end{figure*}

Results for the the 16 qubit IBM device $\mathtt{ibmqx5}$ are shown in Fig. \ref{fig:ibmqx5}. We find that the fuzz peaks at round 2, which is extended to round 4 when the error mitigation is used. Both occur earlier than the peak for randomly chosen pairs for the inverse slices, which occurs at round 5.

The decay of the success rate for MWPM occurs over a similar number of rounds. The main decay continues until around round 4. At this point it begins to converge at around $40\%$, which is the success rate for a randomly guessed pairing.

The error mitigated data decays much more slowly, starting at $100\%$ and maintaining success rates of around $70\%$ as far as round 10. The values are much higher than those for the simulation of randomly chosen pairs for the inverse slices. However, they are still much less than the simulation of correctly chosen pairs with low shots, which still remains above $90\%$ at round 10.

The average difference between the $\tilde{\theta_j}$ and their ideal values is found to be high, and even higher than that for randomly chosen pairs until round 6. Again, little difference is seen between non-mitigated and mitigated data.

\subsection{8 and 19 qubit Rigetti devices}  \label{subsec:rigetti}

\begin{figure*}
	\centering
	\begin{subfigure}[b]{\textwidth}
		\includegraphics[width=\textwidth]{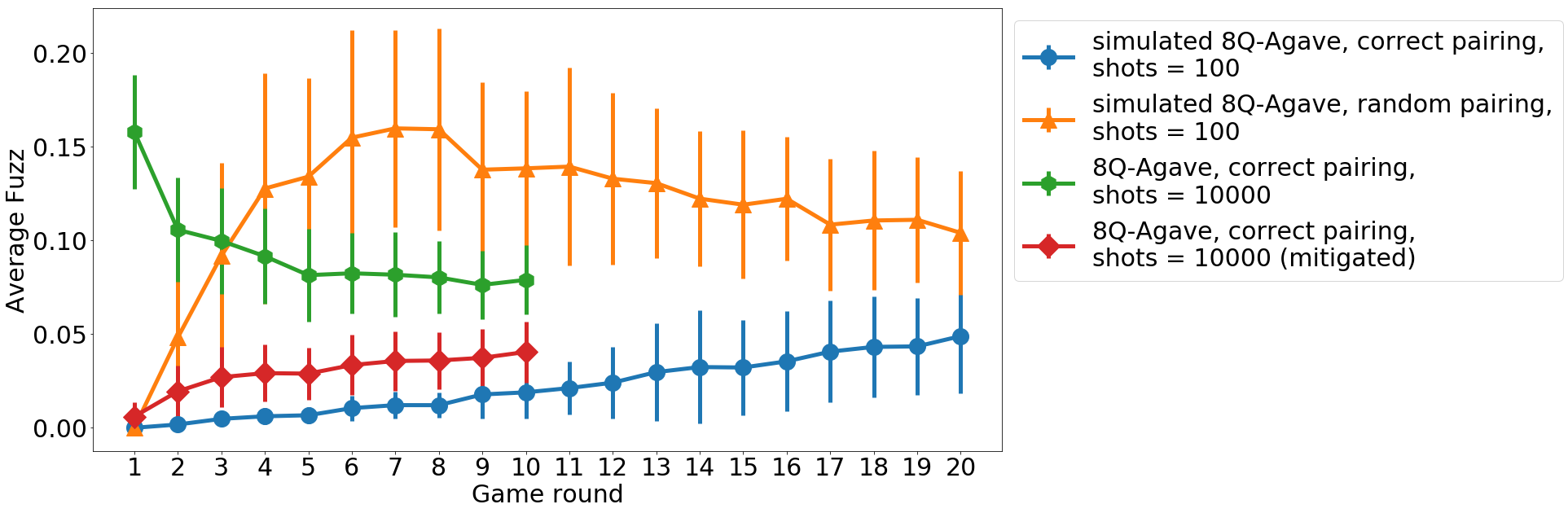}
	\end{subfigure}
	\begin{subfigure}[b]{\textwidth}
		\includegraphics[width=\textwidth]{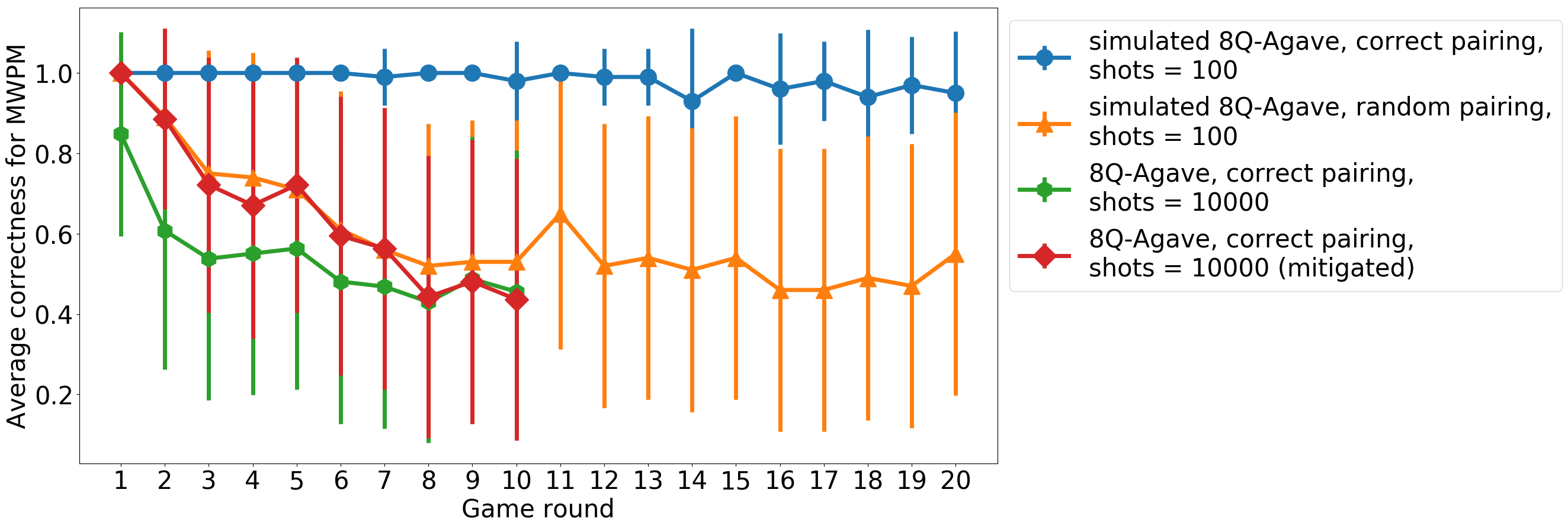}
	\end{subfigure}
	\begin{subfigure}[b]{\textwidth}
		\includegraphics[width=\textwidth]{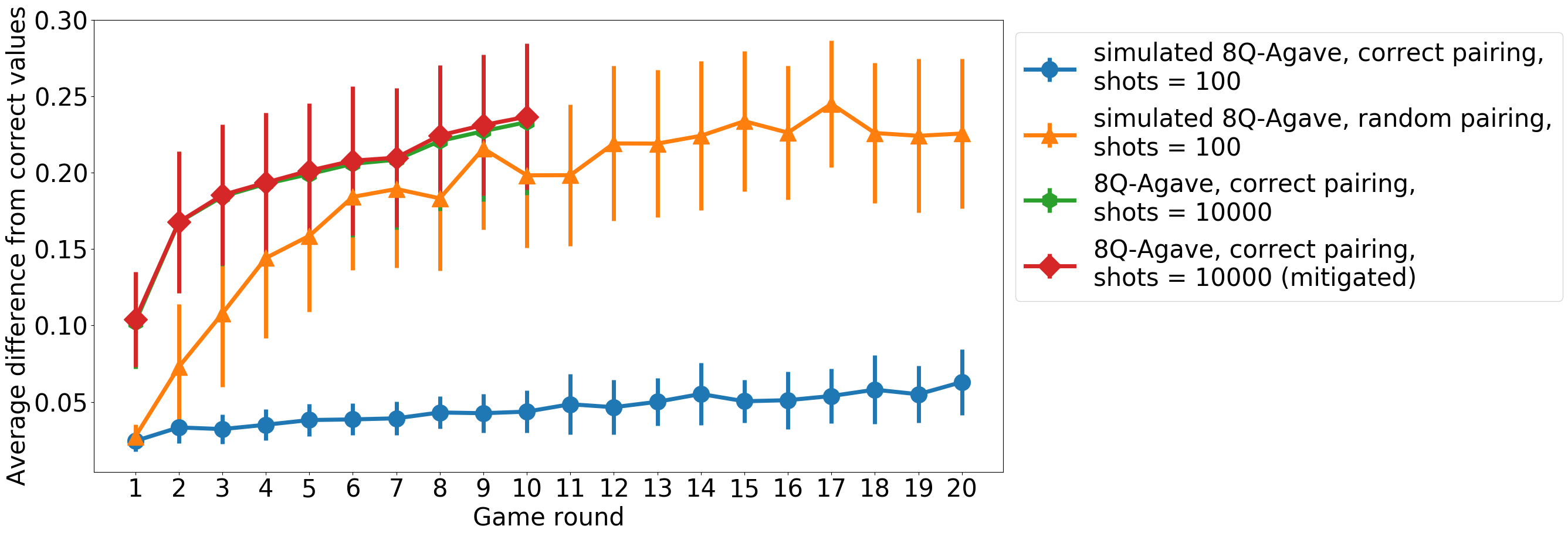}
	\end{subfigure}
	\caption{Results for the Rigetti device $\mathtt{8Q-Agave}$. Each point is the average of 100 samples for simulated data and around 150 samples for the real device. Error bars given by the standard deviation.}\label{fig:agave}
\end{figure*}

Results for the 8 qubit Rigetti device $\mathtt{8Q-Agave}$ are shown in Fig. \ref{fig:agave}. Here we find that the fuzz begins at its peak, and decays thereafter. Error mitigation greatly reduces the values of the fuzz, bring it to values comparable to, but still noticeably higher than, the simulated results for correctly chosen inverse slices.

The success rate for MWPM decays sharply for a number of rounds, before reaching the value for randomly guessed pairs ($50\%$) at around round 6. For error mitigated data, the round 1 value of the success rate achieves $100\%$.

The average difference between the $\tilde{\theta_j}$ and their ideal values is found to be higher than that for randomly chosen pairs for all rounds considered. Again, little difference is seen between non-mitigated and mitigated data.

\begin{figure*}
    \centering
    \begin{subfigure}[b]{\textwidth}
        \includegraphics[width=\textwidth]{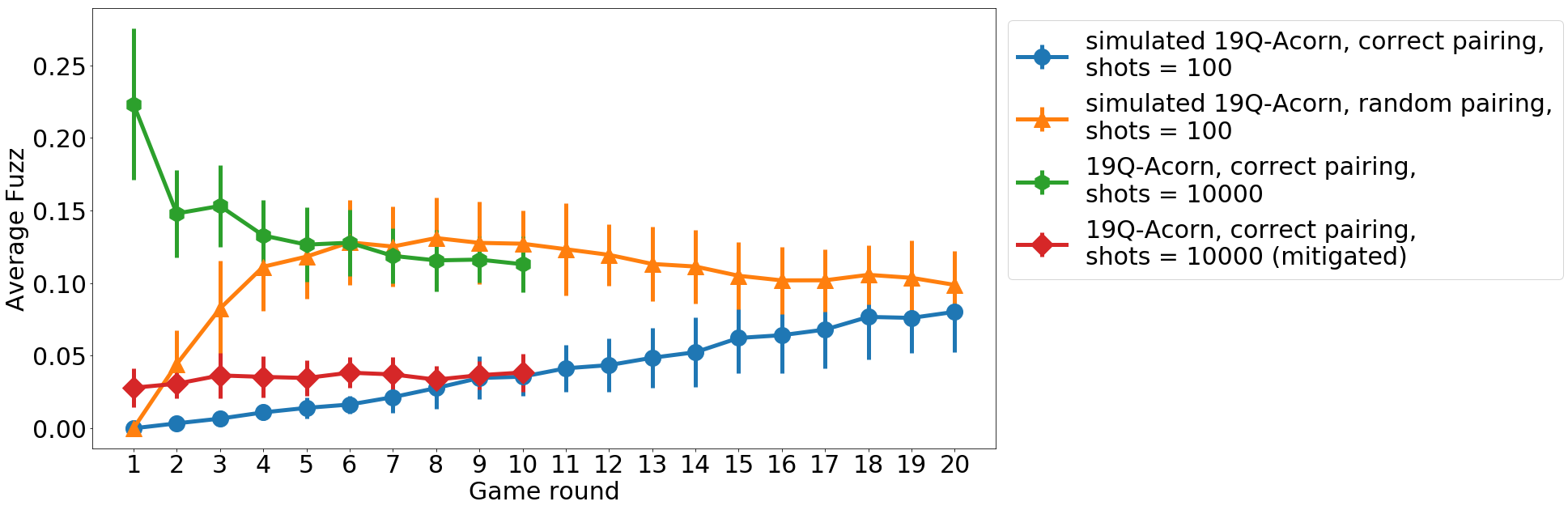}
    \end{subfigure}
    \begin{subfigure}[b]{\textwidth}
        \includegraphics[width=\textwidth]{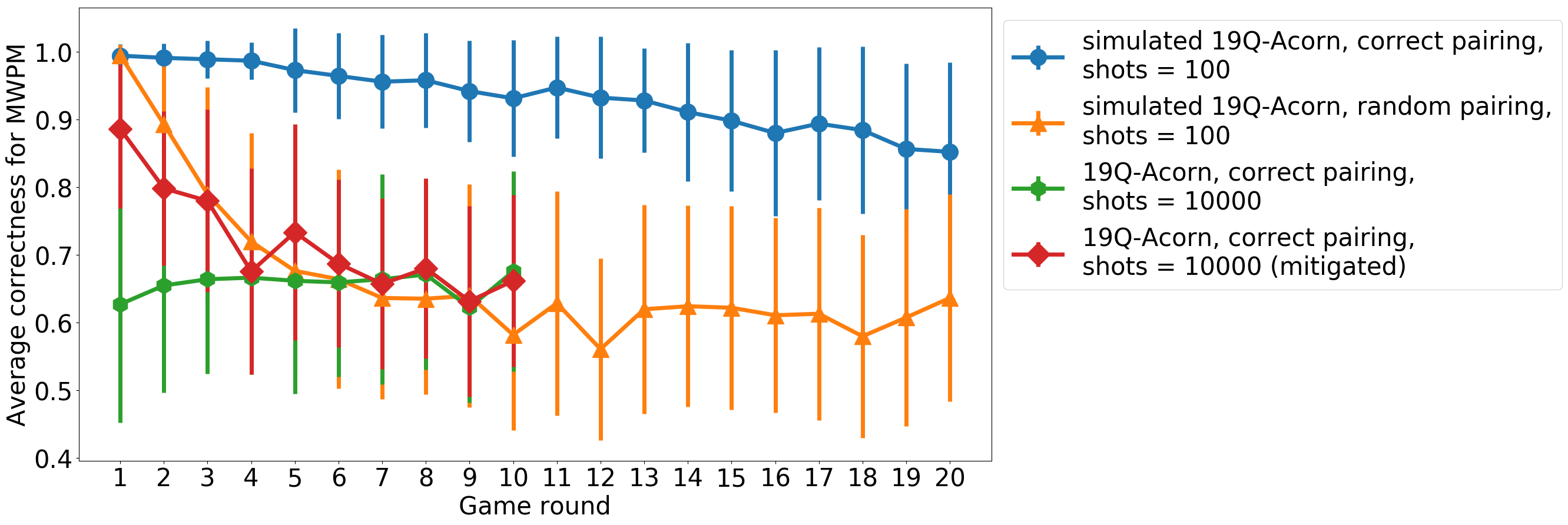}
    \end{subfigure}
    \begin{subfigure}[b]{\textwidth}
        \includegraphics[width=\textwidth]{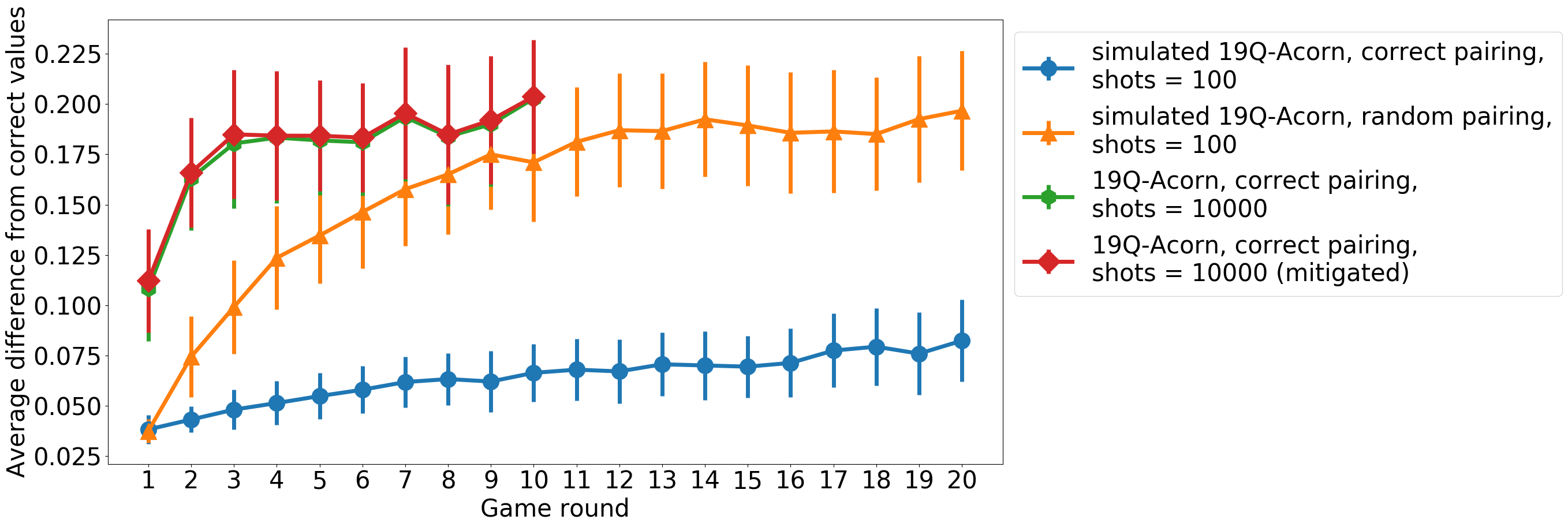}
    \end{subfigure}
    \caption{Results for the Rigetti device $\mathtt{19Q-Acorn}$. Each point is the average of 100 samples for simulated data and around 50 samples for the real device. Error bars given by the standard deviation.}\label{fig:acorn}
\end{figure*}

Results for the the 19 qubit Rigetti device $\mathtt{19Q-Acorn}$ are shown in Fig. \ref{fig:acorn}. The fuzz again begins at its peak and decays thereafter. Error mitigation greatly reduces the values of the fuzz, though it is still difficult to distinguish the point at which the peak occurs. This is partly due to the connectivity of the device, which causes a smooth curve similar to that for a line graph. Nevertheless, it does seem to be delayed until at least round 3 by the mitigation. In either case, it is earlier than the peak for randomly chosen pairs for the inverse slices, which occurs at around round 8.

The success rate for MWPM instantly converges at around $60\%$, which is the success rate for random guessing. For mitigated data, however, the success rate starts as high as round $0.9$ and decays rapidly over the first four rounds.

The average difference between the $\tilde{\theta_j}$ and their ideal values is again found to be higher than that for randomly chosen pairs for all rounds considered, and little difference is seen between non-mitigated and mitigated data.

\subsection{Summary}

To summarize, results from real devices running circuits generated with randomly chosen pairings for the inverse slices were compared to those from simulated instances of both random and correctly chosen pairings. Without error mitigation, the results from the real devices were found to correspond more closely to the simulations of the random pairings, which shows the potent effect of the noise in washing away the expected structure from the outputs. Behaviour much closer to the case of correctly chosen pairings were found when error mitigation was used, especially for the 5 qubit IBM device and with strong effects also for the 16 qubit IBM device and 8 qubit Rigetti device. The strong positive effects of error mitigation show that the devices do indeed have powerful capabilities, but techniques for error mitigation will be very important for unlocking them in the near-term.

\section{Conclusions}

Given the results we have gathered, the conditions required for quantum computational supremacy seem to still be beyond current devices. At the very least, it will require sophisticated methods for error mitigation.

Our results show that current devices certain can support complex programs for non-trivial circuit depth. However, at least for the application considered here, this depth was found to be higher for smaller devices. This highlights the need to not only push for larger devices with better connectivity, but to also ensure that they maintain the quality found for smaller devices.

All data presented in this paper, as well as the software used to gather and process the data, is available at Ref \cite{awesomeness}. Data from other devices will be sought to add from this work, and any contributions of data from other devices will be very welcome.

\section{Acknowledgements}

The author thanks Will Zeng, Alan Ho and Michael Bremner for discussions and Joris van Rantwijk for the use of software for MWPM. This work was supported by the Swiss National Science Foundation and the NCCR QSIT.

Results for this work were generated using hardware from IBM Q and Rigetti, and software from IBM Q (QISKit), Rigetti (Forest) and Project Q. The views expressed are those of the author and do not reflect the official policy or position of any of these entities.

\bibliography{refs}

\pagebreak

\pagebreak

\pagebreak

\pagebreak

\pagebreak

\pagebreak

\pagebreak

\pagebreak

\end{document}